\documentclass[twocolumn,aps,prl,notitlepage,showpacs,floatfix,10pt]{revtex4-1}
\usepackage{epsfig}
\usepackage{graphicx}
\usepackage{dcolumn}
\usepackage{bm}
\usepackage{ulem}
\usepackage{amsfonts,amsmath,amssymb,amstext,dsfont}
\usepackage{mdframed}
\usepackage[percent]{overpic}
\usepackage[colorlinks=true,citecolor=blue]{hyperref}
\hypersetup{colorlinks=true,citecolor=blue,linkcolor=blue,urlcolor=blue}

\usepackage{tikz,bm}
\usetikzlibrary{decorations.pathmorphing}
\usetikzlibrary{decorations.markings}
\usetikzlibrary{decorations.text}
\usetikzlibrary{calc}

\definecolor{purple}{rgb}{0.8,0,0.6}
\definecolor{darkgreen}{rgb}{0.00,0.6,0.00}

\def\mytitle{Acoustogalvanic effect in Dirac and Weyl semimetals}

\begin{document}

\title{\mytitle}
\date{March 26, 2020}

\author{P.~O.~Sukhachov}
\email{pavlo.sukhachov@su.se}
\affiliation{Nordita, KTH Royal Institute of Technology and Stockholm University, Roslagstullsbacken 23, SE-106 91 Stockholm, Sweden}
\author{H. Rostami}
\email{habib.rostami@su.se}
\affiliation{Nordita, KTH Royal Institute of Technology and Stockholm University, Roslagstullsbacken 23, SE-106 91 Stockholm, Sweden}

\begin{abstract}
The acoustogalvanic effect is proposed as a nonlinear mechanism to generate a direct electric current by passing acoustic waves in Dirac and Weyl semimetals. Unlike the standard acoustoelectric effect, which relies on the sound-induced deformation potential and the corresponding electric field, the acoustogalvanic one originates from the pseudo-electromagnetic fields, which are not subject to screening.
The longitudinal acoustogalvanic current scales at least quadratically with the relaxation time, which is in contrast to the photogalvanic current where the scaling is linear.
Because of the interplay of pseudoelectric and pseudomagnetic fields, the current could show a nontrivial dependence on the direction of sound wave propagation. Being within the experimental reach, the effect can be utilized to probe dynamical deformations and corresponding pseudo-electromagnetic fields, which are yet to be experimentally observed in Weyl and Dirac semimetals.
\end{abstract}
\maketitle

{\it Introduction.--}
The investigation of interplay between electric properties and sound waves has a long history and dates back to 1950s~\cite{Parmenter:1953,Akhiezer-Liubarskii:1957,Weinreich-White:1957,Skobov-Kaner:1964,Eckstein:1964} (see also Refs.~\cite{Abrikosov:book-1988,Kittel:book,Gudkov-Gavenda:book}). The generation of electric currents due to sound is known as the {\it acoustoelectric} effect.
Its mechanism is related to a partial uncompensation of sound-induced dynamical deformation potential by electrons in solids.
A sound wave drags charge carriers leading to a measurable current or voltage~\cite{Weinreich-White:1957}.
In low-dimensional systems, surface acoustic waves induced by piezoelectric substrate are routinely used to probe the acoustoelectric response~\cite{Wixforth-Schlapp:1989,Willett-Bishop:1990,Falko-Iordanskii:1993,Hernandez-Minguez-Santos:2018}.
Among them, a valley acoustoelectric effect driven by a surface acoustic wave was recently predicted in two-dimensional (2D) semiconductor~\cite{Kalameitsev-Savenko:2019}. All of the previous proposals can be summarized as follows: sound waves induce an electric field interacting with charge carriers and resulting in an electric current. However, the possibility to generate currents via ``fictitious" strain-induced electromagnetic fields that average to zero over the whole sample was not investigated before.

In recent years, there has been a surge of interest in the fictitious or pseudo-electromagnetic fields in one-, two-, and three-dimensional (3D) strained Dirac materials. As an example, we mention the pseudo-gauge field in carbon nanotubes~\cite{Kane-Mele:1997}, graphene~\cite{Suzuura-Ando:2002,Sasaki-Saito:2005,Katsnelson-Novoselov:2007,Vozmediano-Guinea:2010}, bilayer graphene \cite{Mariani_prb_2012,rostami_prb_2013}, and transition metal dichalcogenides (TMDs)~\cite{cazalilla_prl_2014,rostami_prb_2015}.
A Hall current generated by a time-dependent pseudo-gauge field was also previously discussed in strained Dirac materials~\cite{Vaezi-Vaezi:2013,Oppen-Mariani:2009,Sela-Shalom:2019}. It is worth noting also that the pseudo-gauge fields can be generated in Weyl metamaterials~\cite{Roy-Grushin:2018,Peri-Huber:2019,Jia-Zhang:2019}. In 3D,  the possibility to produce axial gauge fields in strained Weyl semimetals (WSMs) can be noted~\cite{Zhou-Shi:2013,Zubkov:2015,Cortijo-Vozmediano:2015}.
Note that the appearance of strain-induced gauge fields is intimately connected with the fact that quasiparticles in Dirac and Weyl systems are described by the corresponding relativistic-like equations. Conventional metals with a parabolic band dispersion do not allow us to interpret strain in terms of gauge fields.

\begin{figure}[t]
\begin{center}
\includegraphics[width=0.45\textwidth]{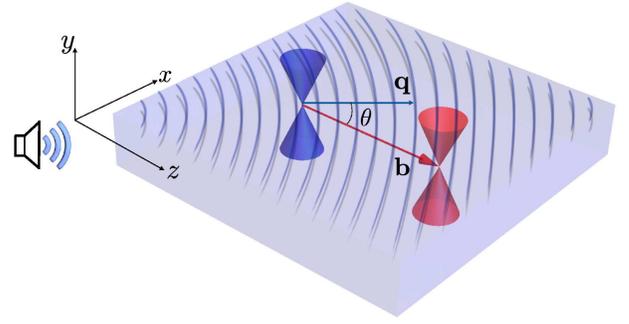}
\end{center}
\caption{
The model setup of a WSM slab where the chiral shift $\mathbf{b}$ is directed along the $z$ axis and the wave vector of the sound wave $\mathbf{q}$ forms an angle $\theta$ with $\mathbf{b}$. Without the loss of generality, we assume that the vectors $\mathbf{q}$ and $\mathbf{b}$ are coplanar.}
\label{fig:setup}
\end{figure}

Dirac semimetals (DSMs) and WSMs represent a special class of solids with relativistic-like quasiparticles~\cite{Wehling-Balatsky:rev-2014,Yan-Felser:2017-Rev,Hasan-Huang:rev-2017,Armitage-Vishwanath:2017-Rev}.
The valence and conduction bands touch at isolated Weyl nodes (Dirac points) allowing one to apply Weyl (Dirac) equations for the description of quasiparticle properties. If the time-reversal ($\mathcal{T}$) symmetry is broken, each Dirac point splits into two Weyl nodes of opposite chiralities separated by the vector $2\mathbf{b}$ (known as the chiral shift~\cite{Gorbar:shift-2009}) in momentum space.
As was shown in Refs.~\cite{Zhou-Shi:2013,Zubkov:2015,Cortijo-Vozmediano:2015}, mechanical strain in WSMs can lead to the generation of the axial gauge field $\mathbf{A}_5$. This field couples to the quasiparticles of opposite chirality as if they have opposite electric charges.
The time-dependent and nonuniform $\mathbf{A}_5({\bf r},t)$ allows for the pseudo-electromagnetic fields $(c{\bf E}_5,\mathbf{B}_5)=(-\partial_t \mathbf{A}_5,\partial_{\bf r}\times\mathbf{A}_5)$. Certain DSM, such as $A_3$Bi ($A=$ Na, K, Rb) and Cd$_3$As$_2$~\cite{Wang:2012,Wang:2013} contain two overlapping copies of $\mathcal{T}$ symmetry broken WSMs with nonzero chiral shifts pointing in opposite directions and also allow for pseudo-electromagnetic fields. (In fact, these DSMs can be classified as $\mathbb{Z}_2$ WSMs~\cite{Gorbar:2014sja}.)
Chirality-selective fields lead to many interesting phenomena~\cite{Ilan-Pikulin:rev-2019}. Among them are the strain-induced chiral magnetic effect and the ``negative" pseudomagnetic resistivity~\cite{Cortijo-Vozmediano:2016,Grushin:2016,Pikulin-Franz:2016,Chernodub-Zubkov:2017,Huang:2017}, quantum oscillations in pseudomagnetic fields~\cite{Pikulin-Franz:2017}, the chiral torsional effect~\cite{Parrikar-Leigh:2014,Sumiyoshi-Fujimoto:2016,Khaidukov-Zubkov:2018}, unusual collective excitations~\cite{Gorbar:2016ygi,Gorbar-Sukhachov:collective,Chernodub-Vozmediano:2019},
axial analogs of the chiral separation and anomalous Hall effects~\cite{Huang:2017}, the lensing of Weyl quasiparticles~\cite{Gorbar:2017dtp,Weststrom-Ojanen:2017,Soto-Garrido-Munoz:2018}, etc. However, to the best of our knowledge, the emergence of direct currents in WSMs and DSMs due to sound-induced {\it dynamical} strain fields was not discussed before.

The study of nonlinear processes such as the photogalvanic (or photovoltaic) effect~\cite{Belinicher_1978,Ivchenko_Pikus_1978,Belinicher_Belinicher_1980}, where a {\it direct} electric current (dc) is generated due to the rectification of driving electromagnetic waves, has recently attracted significant experimental attention in topological materials~\cite{Ma-Gedik:2017,Osterhoudt-Burch-TaAs:2018,Sirica-Taylor-TaAs:2018,Ji-Agarwal-typeII:2019,Ma-Sun-typeII:2019}.
Motivated by these studies, we propose to use dynamical deformations (e.g., sound or acoustic wave) instead of light to generate a dc current. In analogy to the photogalvanic, we dub this phenomenon the {\it acoustogalvanic} effect.
Acoustic waves leads to dynamical local deformations in materials, which are modeled with a propagating displacement vector, ${\bf u} = {\rm Re} [{\bf u}_0 e^{i({\bf q}\cdot{\bf r}-\omega t)}]$, where the sound frequency is $\omega =v_sq$, $\mathbf{q}$ is the wave vector, $v_s$ stands for the sound velocity, and $\mathbf{u}_0$ is the amplitude of the displacement vector. Then, the acoustogalvanic (AG) current is defined as the nonlinear response to the dynamical strain fields
\begin{equation}
\label{intro-j-DC}
j^{\rm dc}_{a} = \chi^{\rm AG}_{abc} u_b u^{*}_c,
\end{equation}
where $ \chi^{\rm AG}_{abc} $ is the acoustogalvanic susceptibility. As we already mentioned above, strains couple as effective oscillating pseudo-electromagnetic fields $\mathbf{E}_5$ and $\mathbf{B}_5$ in WSMs and DSMs. In terms of these fields, the AG current (\ref{intro-j-DC}) can be rewritten as
\begin{equation}
\label{intro-j-DC-1}
j^{\rm dc}_a = \sigma_{abc} E_{5,b} E^\ast_{5,c} +\kappa_{abc} {\rm Re}[E_{5,b} B^\ast_{5,c}] +\gamma_{abc} B_{5,b} B^\ast_{5,c}.
\end{equation}
Note that due to the combined effect of the Berry curvature and pseudo-electromagnetic fields, there will be also alternating currents (ac) in the first order response (see Sec.~S II.A in the Supplemental Material~\cite{SM}). In particular, they are related to the pseudomagnetic analog of the chiral magnetic effect~\cite{Grushin:2016,Huang:2017}. However, due to their alternating nature and a different direction of these currents, they can be easily distinguished from the dc response and will not be considered here.

In what follows, we demonstrate that the pseudo-electromagnetic fields lead to a nontrivial AG response of WSMs and DSMs, where a dc current is generated in second order processes. By using the chiral kinetic theory~\cite{Son:2013,Stephanov:2012,Son-Yamamoto:2012,Son-Spivak:2013,Gorbar:2016ygi} as well as applying longitudinal sound waves, we calculate the intra-band contribution to the AG current in a doped $\mathcal{T}$ symmetry-broken WSMs and certain DSMs~\cite{comment1}. As for possible material realizations of the present setup, we mention the WSM EuCd$_2$As$_2$~\cite{Wang-Canfield-EuCd2As2:2019,Ma-Shi:2019} and the DSMs $A_3$Bi ($A=$ Na, K, Rb) and Cd$_3$As$_2$~\cite{Wang:2012,Wang:2013}.
The origin of strain-induced electric currents is related to the {\it acoustoelectric drag effect}, where the AG current vanishes when the wave vector of sound wave goes to zero $q\to 0$.
In addition, an interplay of the strain-induced pseudoelectric and pseudomagnetic fields allows for a nontrivial dependence of the AG current on the direction of sound wave propagation. Finally, we provide estimations of the proposed acoustogalvanic effect.
As for the practical implications, we believe that it can be useful for investigating dynamical deformations. Finally, while we concentrate on the case of rectified electric current, the dc chiral current is also possible and is discussed in the Supplemental Material~\cite{SM}.

{\it Model.--}
An effective Hamiltonian of strained WSMs in the vicinity of Weyl nodes is given by
\begin{equation}
\label{model-H-def}
{\cal H}_\lambda = \lambda v_{\rm F} {\bm \sigma} \cdot \left[{\bf p} + \frac{e}{c} \lambda {\bf A}_5({\bf r},t)\right]+ D({\bf r},t),
\end{equation}
where $\lambda=\pm$ is chirality, ${\bm \sigma}=(\sigma_x,\sigma_y,\sigma_z)$ is the vector of Pauli matrices, ${\bf p} \equiv -i \hbar \partial_{\bf r}$ is the momentum, ${\bf A}_5$ is the strain-induced gauge field, and $D$ is the deformation potential.
It is easy to verify that the dispersion relation of the undeformed system in the vicinity of Weyl nodes is linear, $\epsilon^{(0)}_{\eta,{\bf p}} = \eta v_{\rm F} p$, where $\eta=+$ and $\eta=-$ correspond to the conduction and valence bands, respectively.

For definiteness, we consider a longitudinal sound wave, i.e., ${\bf u}_0 = u_0\hat{\bf q}$, and, without the loss of generality, set ${\bf  b} = b \hat{\bf z}$. The corresponding model setup is presented in Fig.~\ref{fig:setup}. The sound-induced deformation leads to the following axial gauge field~\cite{Cortijo-Vozmediano:2015,Cortijo-Vozmediano:2016}
\begin{align}
\label{model-A5-def}
&A_{5,i}= -\frac{c\hbar b}{e} \Big[ \beta  u_{iz} +\tilde{\beta}(b) \delta_{iz}\sum_{j}u_{jj}\Big].
\end{align}
The deformation potential $D$ can be represented as the series in displacement field $D=\sum_n D^{(n)}$. For instance, the deformation potential in the first order $n=1$ is $D^{(1)} \propto \sum_{j} u_{jj}$. Note that $u_{ij} = \left(\partial_{i} u_j+\partial_{j} u_i\right)/2$ is the linearized strain tensor as well as $\beta$ and $\tilde{\beta}(b)$ are related to the Gr\"{u}neisen parameters. Similarly to graphene~\cite{Vozmediano-Guinea:2010}, we assumed that the deformation potential $D$ is isotropic and momentum independent.
The electric current $\bf j$ and charge $\rho$ densities are defined as~\cite{Xiao-Niu:rev-2010,Son-Yamamoto:2012,Son:2013}
\begin{align}
\label{Model-sound-current-j}
\mathbf{j} &= -e\sum_{\lambda,\eta=\pm} \sum_{\bf p} 
\eta \Big[ (\partial_t\mathbf{r}) L_{\eta,\lambda} f_{\eta,\lambda}
+\partial_{\bf r}\times
(\epsilon_{\eta,\mathbf{p}}\mathbf{\Omega}_{\eta,\lambda}f_{\eta,\lambda})
\Big],
\nonumber\\
\rho &= -e\sum_{\lambda,\eta=\pm} \sum_{\bf p}
\eta L_{\eta,\lambda} f_{\eta,\lambda}
\end{align}
with $\sum_{\bf p}\equiv \int d^3p/(2\pi\hbar)^3$.
Note that $L_{\eta,\lambda}=1- e\lambda\left(\mathbf{B}_{5}\cdot\mathbf{\Omega}_{\eta,\lambda}\right)/c$ stands for the phase-space volume, which is renormalized by the Berry curvature ${\bm \Omega}_{\eta,\lambda}=\sum_n {\bm \Omega}^{(n)}_{\eta,\lambda}$. In undeformed systems, the latter has a monopole-like structure ${\bm \Omega}^{(0)}_{\eta,\lambda} =\lambda \eta \hat{\bf p}/(2 p^2)$  (see Sec.~S I. in the Supplemental Material~\cite{SM} for the fields-induced corrections to the Berry curvature). The last term in the electric current ${\bf j}$ corresponds to the orbital magnetization ($\propto \partial_{\bf r}\times{\bf M}$) (see, e.g., Ref.~\cite{Xiao-Niu:rev-2010}).
The distribution function $f_{\eta,\lambda}$ for quasiparticles of each chirality is obtained by solving the Boltzmann equation in the presence of both pseudoelectric and pseudomagnetic fields
\begin{align}
\partial_t f_{\eta,\lambda} +\left(\partial_t\mathbf{p}\right)\cdot \partial_\mathbf{p} f_{\eta,\lambda} +\left(\partial_t\mathbf{r}\right)\cdot \partial_{\bf r} f_{\eta,\lambda}=-\frac{f_{\eta,\lambda}-f_{\eta,\lambda}^{(0)}}{\tau}.
\label{Model-sound-kinetic-equation}
\end{align}
Here $f^{(0)}_{\eta,\lambda} = 1/\left[e^{\eta\left(\epsilon_{\eta,\mathbf{p}}+D-\mu_\lambda\right)/T}+1\right]$ is the local equilibrium distribution function where the divergent vacuum contribution was subtracted, $T$ is temperature in the energy units, and $\mu_\lambda$ is the chemical potential, which contains corrections from displacement fields, i.e., $\mu_\lambda =\sum_n\mu^{(n)}_\lambda$ where $\mu^{(0)}=\mu$. It can be decomposed into electric $\mu^{(n)}$ and chiral $\mu^{(n)}_5$ parts: $\mu^{(n)}_\lambda=\mu^{(n)} +\lambda \mu^{(n)}_5$.
Note also that the quasiparticle dispersion obtains additional field-induced corrections $\epsilon_{\eta,\mathbf{p}}=\sum_n\epsilon^{(n)}_{\eta,\mathbf{p}}$ in the presence of pseudo-electromagnetic fields~\cite{Gao-Niu:2014,Gao-Niu:2015} (for an explicit expression, see Sec.~S I. in the Supplemental Material~\cite{SM}).
For simplicity, we utilized a simple relaxation time approximation for the collision integral, where $\tau$ is the intra-node relaxation time and the inter-node processes were neglected. The equations of motion for the chiral quasiparticles are strongly modified by the Berry curvature and read as~\cite{Xiao-Niu:rev-2010}
\begin{align}
\label{Model-sound-kinetic-r-dot}
\partial_t\mathbf{r} &= \frac{\mathbf{v}_{\eta,\mathbf{p}} -e\big(\tilde{\mathbf{E}}_{\lambda}\times\mathbf{\Omega}_{\eta,\lambda}\big)
-\lambda \frac{e}{c}\big(\mathbf{v}_{\eta,\mathbf{p}}\cdot\mathbf{\Omega}_{\eta,\lambda}\big)\mathbf{B}_{5}}{L_{\eta,\lambda}},
\\
\label{Model-sound-kinetic-p-dot}
\partial_t\mathbf{p} &= -\frac{e\tilde{\mathbf{E}}_{\lambda} +\lambda\frac{e}{c}\big(\mathbf{v}_{\eta,\mathbf{p}}\times \mathbf{B}_{5}\big)}{L_{\eta,\lambda}},
\end{align}
where $\mathbf{v}_{\eta,\mathbf{p}}= \partial_{\bf p} \epsilon_{\eta,{\bf p}}$ stands for the quasiparticle velocity. Note also that the effective pseudoelectric field is $\tilde{\mathbf{E}}_{\lambda} = \lambda\mathbf{E}_5 +\partial_{\bf r}\left(D-\mu_{\lambda}\right)/e$ and we took into account that for a longitudinal sound wave $\tilde{\mathbf{E}}_5\cdot\mathbf{B}_5=0$.

Having defined the key aspects of the model, let us discuss how to calculate the current density. We assume that the deformations are sufficiently weak to allow for a perturbative solution to the Boltzmann equation~(\ref{Model-sound-kinetic-equation}). Since the propagation of sound distorts the ionic lattice, electrons tend to compensate local deviations from the charge neutrality and modify local electric chemical potential.
As follows from the continuity relation for an electric current, the deviations of electric chemical potential $\mu^{(n)}$ compensate the deformation potential $D^{(n)}$. For the model at hand, the compensation is exact in the first order $n=1$.
Then, the continuity relation for a chiral current allows for residual corrections to the chiral chemical potential that renormalize the effective electric field as $\tilde {\bf E}_\lambda \to \lambda \tilde{\bf E}_5$ where $\tilde{\bf E}_5 ={\bf E}_5 - \partial_{\bf r}\mu_5^{(1)}/e$ at $n=1$ (see also Sec.~II.A in the Supplemental Material~\cite{SM}).
The corresponding chiral chemical potential at $\omega\tau\ll1$ reads as
\begin{align}
\label{Model-mu5}
\mu^{(1)}_5 \approx -\frac{e\tau v^2_{\rm F} \left(1+i\omega\tau\right) }{3\omega} \left(\mathbf{E}_5\cdot\mathbf{q}\right) e^{i({\bf q}\cdot {\bf r}-\omega t)} + c.c.
\end{align}

{\it Acoustogalvanic response.--}
In the case of an arbitrary direction of sound wave propagation, both pseudoelectric $\mathbf{E}_5$ and pseudomagnetic $\mathbf{B}_5$ fields are generated. While the general expressions for the response tensors $\sigma_{abc}$, $\kappa_{abc}$, and $\gamma_{abc}$ are given in Sec.~S~II.B in the Supplemental Material~\cite{SM}, here we focus on two limiting cases of sound wave propagation with respect to the chiral shift: ({\it i}) $\mathbf{q}\parallel \mathbf{b}$ and ({\it ii}) $\mathbf{q}\perp \mathbf{b}$. They correspond to $\theta=0$ and $\theta=\pi/2$, respectively, in Fig.~\ref{fig:setup}.
\begin{figure}[t]
\begin{center}
\includegraphics[width=0.365\textwidth]{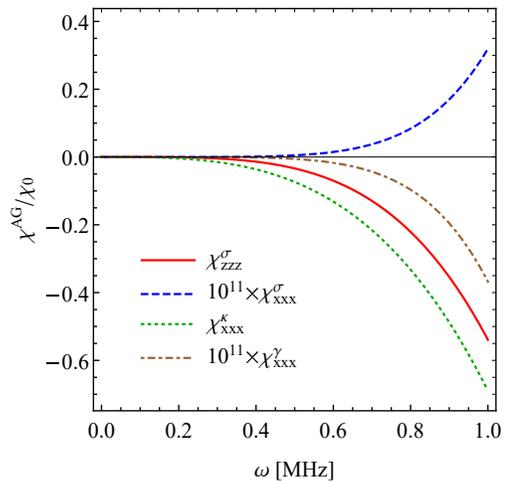}
\end{center}
\caption{The dependency of the acoustogalvanic susceptibility $\chi^{\rm AG}$ on the sound frequency $\omega$ for $\mathbf{q}\parallel \mathbf{b}$ ($\theta=0$) (solid red curve) and $\mathbf{q}\perp \mathbf{b}$ ($\theta=\pi/2$) (the other three curves). Here $\chi_0 = 10^{4}~{\rm A/cm^4}$.}
\label{fig:cc-II-jz-DC}
\end{figure}

Case ({\it i}): Let us start with the case of the second-order response at $\mathbf{q}\parallel\mathbf{b}$. As is easy to verify by using Eq.~(\ref{model-A5-def}), the pseudomagnetic field $\mathbf{B}_5$ is absent in this case and the pseudoelectric one $\mathbf{E}_5$ is directed along $\mathbf{q}$.
The only relevant element of the nonlinear conductivity is $\sigma_{zzz}$. In the leading order in small $\omega \tau$ and large $v_{\rm F}/v_s$ it reads as~\cite{SM}
\begin{equation}
\label{cc-second-j-DC-simpl}
\sigma_{zzz}\approx - \frac{e^3}{\hbar^2} \frac{v_{\rm F}}{v_s} \frac{ \mu \tau^2}{18\pi^2 \hbar}  + {\cal O}[(\omega\tau)^2].
\end{equation}
Intriguingly, we find that, due to the contribution of the chiral chemical potential (\ref{Model-mu5}), the nonlinear conductivity here scales as $\tau^2$ and it is independent of $\omega$. Such a dependence on frequency is clearly different from that for the conventional optical rectification, where $\sigma^{\rm PG} \propto \tau$ or $1/\omega$~\cite{Sipe_prb_2000,Juan_nc_2017,Rostami_prb_2018, Matsyshyn_arxiv_2019,Juan_arxiv_2019}.
The corresponding component of the acoustogalvanic response function $\chi^{\rm AG}_{zzz}$ follows~\cite{SM}
\begin{equation}
\label{cc-second-chi-AG-zzz}
\chi^{\rm AG}_{zzz} = \frac{\omega^4}{v^2_s} \frac{\hbar^2 b^2}{e^2}
[\beta+ \tilde\beta(b)]^2  \sigma_{zzz}.
\end{equation}
As one can see, $\chi^{\rm AG}_{zzz}$ grows with the sound frequency as $\omega^4$ owing to quadratic dependence of the pseudoelectric field, $E_5 \propto  \omega^2$.
Such a strong frequency dependence is one of the characteristic features of the acoustogalvanic response.

Case ({\it ii}): In the case $\mathbf{q}\perp \mathbf{b}$ (without the loss of generality, $\mathbf{q}\parallel\hat{\mathbf{x}}$), both $\mathbf{E}_5$ and $\mathbf{B}_5$ are nonzero, which enriches the dynamics of the system. In the leading order in $\omega \tau$, the following components of the response tensors are relevant~\cite{SM}
\begin{align}
\label{cc-second-sigma-perp}
\sigma_{xzz}&\approx \frac{e^3}{\hbar^2} \frac{v_{\rm F}}{v_s} \frac{ \mu\tau^2}{30\pi^2 \hbar }  (\omega\tau)^2 + {\cal O}[(\omega\tau)^3]~,\\
\label{cc-second-kappa-perp}
\kappa_{xzy}&\approx \frac{e^3}{\hbar^2} \frac{v_{\rm F}}{c} \frac{ \mu \tau^2}{12\pi^2 \hbar} +G_1(\mu, T,\Lambda_{\rm IR}) + {\cal O}[(\omega\tau)^2],
\\
\label{cc-second-gamma-perp}
\gamma_{xyy}&\approx (\omega\tau)^2 G_2(\mu, T) + {\cal O}[(\omega\tau)^3].
\end{align}
The explicit definitions of $G_1(\mu, T,\Lambda_{\rm IR})$ and $G_2(\mu, T)$ are given in Eqs.~(S71) and (S72) in the Supplemental Material~\cite{SM}. Note that, in the second-order chiral kinetic theory, function $G_1(\mu, T,\Lambda_{\rm IR})$ depends on the regularization scheme. While we used a simple infrared cutoff $\Lambda_{\rm IR}$, a more refined treatment might be required (see also the discussion in Sec~S~II.B in the Supplemental Material~\cite{SM}). As one can see, the leading order contribution in $\omega\tau \ll 1$ stems from the interplay between the pseudoelectric and pseudomagnetic fields quantified by $\kappa_{xzy}$ in Eq.~(\ref{cc-second-kappa-perp}). Unlike the response to the pseudoelectric fields, where, as in the case ({\it i}), electric current is also directed along the chiral shift, $\kappa_{xzy}$ is related to the Hall-like response $\propto\mathbf{E}_5\times \mathbf{B}_5$. Note also that this part of the response contains terms insensitive to the relaxation time.

The corresponding acoustogalvanic susceptibility contains three contributions $\chi^{\rm AG}_{xxx}= \chi^{\sigma}_{xxx}+\chi^{\kappa}_{xxx}+\chi^{\gamma}_{xxx}$. Explicit expressions for these terms are given in Eqs.~(S68)--(S70) in the Supplemental Material~\cite{SM}. Their numerical values are depicted in Fig.~\ref{fig:cc-II-jz-DC}. The difference in magnitude between $\chi^{\sigma}_{xxx}$ and $\chi^{\gamma}_{xxx}$ as well as $\chi^{\kappa}_{xxx}$
is related to the fact that the relative scaling of the former with respect to the latter is $(\omega \tau)^2$ for small $\omega \tau\ll1$. In addition, the strong frequency dependence of the acoustogalvanic susceptibility is clearly evident from the figure.
Furthermore, we present the dependence of the current components $j_x^{\rm dc}$ and $j_z^{\rm dc}$ on the angle between the sound wave vector and the chiral shift in Fig.~\ref{fig:cc-II-jz-DC-theta}. While the angular profile of the former component is $\sim \sin{\theta} \left[1+A_0 \cos{(2\theta)}\right]$, $j_z^{\rm dc} \propto\cos{\theta}\left[1-A_0 \cos{(2\theta)}\right]$, where $A_0$ is a combination of functions $G_1(\mu, T,\Lambda_{\rm IR})$, $G_2(\mu, T)$, and terms $\propto\mu$. The terms with $\cos{(2\theta)}$ cause a nontrivial modulation observed in Fig.~\ref{fig:cc-II-jz-DC-theta}. In particular, due to the interplay of the pseudoelectric and pseudomagnetic fields, $j_x^{\rm dc}$ attains its maximal values at $\theta\approx \pi/4$ and has a characteristic butterfly-like angular profile.
For our numerical estimates, we used the numerical parameters valid for the DSM Cd$_3$As$_2$~\cite{Freyland-Madelung:book,Wang-Yamazaki:2007,Neupane-Hasan-Cd3As2:2014,Liu-Chen-Cd3As2:2014,Li-Yu-Cd3As2:2015}: $v_{\rm F}\approx 1.5\times 10^8~{\rm cm/s}$, $\mu\approx 200~{\rm meV}$, $b\approx 1.6~{\rm nm}^{-1}$, $ v_s \approx 2.3\times 10^5~{\rm cm/s}$, and $\tau\approx~1~{\rm ps}$.
In addition, we assume $\beta\approx 1$, $T=5~{\rm K}$, and $\tilde{\beta}(b)\approx 1$.

\begin{figure}[t]
\begin{center}
\includegraphics[width=0.33\textwidth]{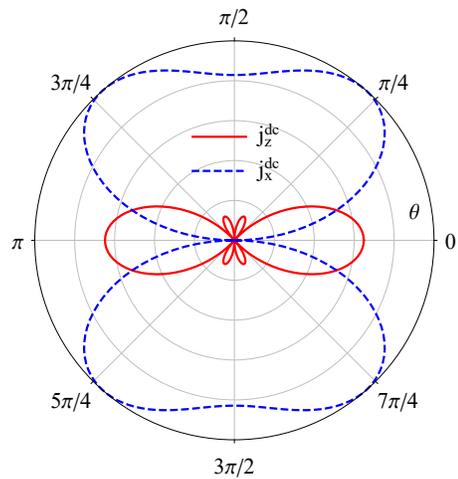}
\end{center}
\caption{The dependency of the rectified current components $j_z^{\rm dc}$ and $j_x^{\rm dc}$ on the angle $\theta$ between the chiral shift $\mathbf{b}$ and the wave vector $\mathbf{q}$. We fixed $\omega=1~\mbox{MHz}$. A typical value of the rectified current is
$j^{\rm dc}\approx 54~u_0^2/{\rm \mu m^2}~[{\rm \mu A/{\rm cm}^2]}$ at $\theta=0$. }
\label{fig:cc-II-jz-DC-theta}
\end{figure}

{\it Discussion and summary.--}
In this study, a nonlinear mechanism to generate a rectified electric current by passing sound waves in WSMs and certain DSMs is proposed. Unlike the conventional acoustoelectric effect, the dc current is produced by the strain-induced pseudo-electromagnetic fields rather than real electric fields. Therefore, in analogy to the photogalvanic effect, we called this mechanism the acoustogalvanic rectification.

The sound-induced dc current quickly grows with a sound frequency. This profound difference to the usual optical rectification is explained by the fact that the pseudo-electromagnetic fields are determined by the dynamics of the deformation vector and, therefore, grow with frequency. By using the realistic model parameters, we estimated that the acoustogalvanic current should be experimentally observable for high frequencies (e.g., ultrasound) and amplitudes of sound. Indeed, the order of magnitude of the AG current is $I^{\rm dc}\sim 100~{\rm nA}$ for $\omega=10~{\rm MHz}$, $u_0=10~{\rm nm}$, and mm-sized crystals. For example, photocurrents of such magnitudes were recently observed in, e.g., Ref.~\cite{Ma-Sun-typeII:2019}. Such a current is also comparable to that in TMDs, where $I^{\rm dc}\sim 10\,\mu \mbox{A}$ for GHz frequencies~\cite{Kalameitsev-Savenko:2019}, and in graphene, where $I^{\rm dc}\sim 0.1 -1\,\mu \mbox{A}$ for MHz frequencies~\cite{Hernandez-Minguez-Santos:2018}. As an additional advantage over the acoustoelectric effect, which relies on an electric field, pseudo-electromagnetic fields that drive acoustogalvanic currents are not subject to screening and could attain significantly high values.

Let us briefly comment on the case of DSMs such as Na$_3$Bi and Cd$_3$As$_2$ where both $\mathcal{T}$ and parity ($\mathcal{P}$) symmetries are preserved. In each copy of WSMs that constitute these DSMs, the direction of pseudo-electromagnetic fields is opposite and cancels the majority of first-order effects such as, e.g., the anomalous Hall effect. However, the second-order response will be doubled with respect to simple $\mathcal{T}$ symmetry broken WSMs.

Further, we discuss the AG effect when $\mathcal{T}$ symmetry is preserved but $\mathcal{P}$ symmetry is broken. By using a simple model where the non-degenerate Weyl nodes are separated only in energy by $2b_0$, it can be shown that there will be the scalar component of the gauge field $A_{0,5}$~\cite{SM}.
The direction of the corresponding pseudoelectric field $\mathbf{E}_5 = -\partial_{\mathbf{r}}A_{0,5}$ is determined exclusively by the sound wave vector. Then, the susceptibility tensor is isotropic $\chi^{\rm AG}_{abc}\sim -\delta_{ab}\delta_{bc}b_0^2v_F \mu \tau^2 \omega^4 /v_s^5$ and can be comparable in magnitude to that in Fig.~\ref{fig:cc-II-jz-DC}.
In the case of WSMs with multiple pairs of Weyl nodes, such as transition metal monopnictides (e.g., TaAs), AG currents should be generated independently for each pair. Therefore, the rectified current $\mathbf{j}^{\rm dc}$ should be present regardless of the direction of the wave vector $\mathbf{q}$.
In addition, the acoustogalvanic effect should occur in type-II WSMs~\cite{Soluyanov:2015}, as long as strains can be interpreted in terms of axial gauge fields~\cite{SM}. Thus, the existence of the acoustogalvanic effect relies primarily on whether strains can be interpreted as axial gauge fields rather than specific properties of materials.

Finally, while in this study we concentrated on the case of 3D WSMs, we believe that our qualitative results can be applied for 2D Dirac materials such as graphene and TMDs. Indeed, since dynamical strain also generates pseudoelectric fields in these materials, one can follow the same steps in the calculation of acoustogalvanic response as discussed in our study. Thus, a dc current could be also generated in 2D materials due to the acoustic drag effect.

\let\oldaddcontentsline\addcontentsline
\renewcommand{\addcontentsline}[3]{}
\begin{acknowledgments}
{\it Acknowledgements.--}
We are grateful to A.~V.~Balatsky for useful discussions. P.O.S. thanks E.~V.~Gorbar for critical comments.
We acknowledge the support from the VILLUM FONDEN via the Centre of Excellence for Dirac Materials (Grant No.~11744), the European Research Council under the European Unions Seventh Framework ERS-2018-SYG 810451 HERO, and the Knut and Alice Wallenberg Foundation KAW 2018.0104. H.R. acknowledges support from the Swedish Research Council (VR 2018-04252).
\end{acknowledgments}
\let\addcontentsline\oldaddcontentsline

\let\oldaddcontentsline\addcontentsline
\renewcommand{\addcontentsline}[3]{}

\let\addcontentsline\oldaddcontentsline

\newpage
\clearpage
\widetext
\begin{center}
\textbf{\large \mytitle~--- Supplemental Material}
\vspace{0.5cm}
P.~O.~Sukhachov,$^{1}$
and H. Rostami$^{1}$\\[4pt]
$^{1}${\small\it Nordita, KTH Royal Institute of Technology and Stockholm University, Roslagstullsbacken 23, SE-106 91 Stockholm, Sweden}
\end{center}
\setcounter{equation}{0}
\setcounter{figure}{0}
\setcounter{table}{0}
\setcounter{page}{1}
\makeatletter
\renewcommand{\thepage}{S\arabic{page}}
\renewcommand{\theequation}{S\arabic{equation}}
\renewcommand{\thesection}{S \MakeUppercase{\roman{section}}}
\renewcommand{\thefigure}{S\arabic{figure}}
\renewcommand{\bibnumfmt}[1]{[S#1]}
\renewcommand{\citenumfont}[1]{S#1}

\maketitle
\tableofcontents

\section{S I. Chiral kinetic theory}
\label{sec:Model-general}

In this section, we present the key details of the chiral kinetic theory (CKT) valid up to the second order in (pseudo)electromagnetic fields. The Boltzmann equation of the latter reads as~
[S1--S5]
\begin{eqnarray}
&&\partial_t f_{\lambda} +\frac{1}{1-\frac{e}{c}\left(\mathbf{B}_{\lambda}\cdot\mathbf{\Omega}_{\lambda}\right)}
\Bigg\{\Big[]-e\tilde{\mathbf{E}}_{\lambda} -\frac{e}{c}\left(\mathbf{v}_{\mathbf{p}}\times \mathbf{B}_{\lambda}\right) +\frac{e^2}{c}\left(\tilde{\mathbf{E}}_{\lambda}\cdot\mathbf{B}_{\lambda}\right)\mathbf{\Omega}_{\lambda}\Big]\cdot
\partial_\mathbf{p} f_{\lambda} \nonumber\\
&&+\Big[\mathbf{v}_{\mathbf{p}} -e\left(\tilde{\mathbf{E}}_{\lambda}\times\mathbf{\Omega}_{\lambda}\right)
-\frac{e}{c}\left(\mathbf{v}_{\mathbf{p}}\cdot\mathbf{\Omega}_{\lambda}\right)\mathbf{B}_{\lambda}\Big]\cdot \partial_{\mathbf{r}} f_{\lambda}\Bigg\}=I_{\rm coll}(f_{\lambda}),
\label{Model-CKT-kinetic-equation}
\end{eqnarray}
where $\tilde{\mathbf{E}}_{\lambda} = \mathbf{E}_{\lambda}  +(1/e)\partial_{\mathbf{r}} \epsilon_{\mathbf{p}}$ is the effective electric field with the last term corresponding to the deformation-induced change in the quasiparticle energy dispersion $\epsilon_{\mathbf{p}}$, $f_{\lambda}$ is the distribution function of quasiparticles of a given chirality $\lambda=\pm$, $-e$ is the charge of an electron, and $c$ is the speed of light.
For the sake of simplicity, we drop the subscript $\eta$ used in the main text, which stands for the band index.
Further, $\mathbf{E}_{\lambda} = \mathbf{E} +\lambda \mathbf{E}_5$ and $\mathbf{B}_{\lambda} = \mathbf{B} +\lambda \mathbf{B}_5$ are effective electric and magnetic fields containing electromagnetic $\mathbf{E}$ and $\mathbf{B}$ as well as pseudo-electromagnetic $\mathbf{E}_5$ and $\mathbf{B}_5$ fields, $\mathbf{\Omega}_{\lambda}$ is the Berry curvature monopole, and $\mathbf{v}_{\mathbf{p}}= \partial_\mathbf{p}\epsilon_{\mathbf{p}}$ is the quasiparticle velocity. The {\it global equilibrium} distribution (Fermi--Dirac) function is
\begin{equation}
f^{\rm eq}_{\lambda} = \frac{1}{e^{\left(\epsilon_{\mathbf{p}}^{(0)} -\mu_{\lambda}^{(0)}\right)/T}+1}.
\label{Model-CKT-equilibrium-function}
\end{equation}
Here $\mu_{\lambda}^{(0)}=\mu^{(0)}+\lambda\mu_5^{(0)}$ is the effective chemical potential for the right- ($\lambda=+$) and left-handed ($\lambda=-$) quasiparticles, $\mu^{(0)}$ is the electric chemical potential, $\mu_5^{(0)}$ is the chiral chemical potential, $T$ is temperature in the energy units, $\epsilon_{\mathbf{p}}^{(0)}=\eta v_Fp$ is the dispersion relation, $v_F$ is the Fermi velocity, and $\eta=\pm$ corresponds to the conduction ($\eta=+$) and valence ($\eta=-$) bands, respectively. Superscript $(0)$ stands for the undeformed system.
Finally, the collision integral in the Boltzmann equation (\ref{Model-CKT-kinetic-equation}) is denoted by $I_{\rm coll}(f_{\lambda})$. In the relaxation time approximation it is given by $I^{\rm intra}_{\rm coll}=-\left(f_{\lambda}-f^{(0)}_{\lambda}\right)/\tau$, where $\tau$ is the intra-node relaxation time and $f^{(0)}_{\lambda}$ is the {\it local equilibrium} function, which will be defined in Eq.~(\ref{app-cc-dp-SK-screen-f}). Since the inter-node relaxation usually require large momentum transfer, the corresponding relaxation time $\tau_5$ is much longer than that for intra-node processes. Therefore, for simplicity, we will neglect inter-node processes. It is convenient to separate the contribution of filled states in the distribution function (\ref{Model-CKT-equilibrium-function}) for the hole band ($\eta=-$), i.e.,
\begin{equation}
f^{\rm eq}_{\lambda} =\delta_{\eta,+}\frac{1}{e^{\left(\epsilon_{\mathbf{p}}^{(0)}-\mu_{\lambda}^{(0)}\right)/T}+1}+ \delta_{\eta,-} \left[1-\frac{1}{e^{-\left(\epsilon_{\mathbf{p}}^{(0)}-\mu_{\lambda}^{(0)}\right)/T}+1}\right].
\label{Model-CKT-equilibrium-function-eta}
\end{equation}
In what follows, the contribution of the filled states $\delta_{\eta,-}$ will be subtracted leading to the multiplier $\eta$ in the expressions for the charge and current densities (\ref{Model-CKT-charge}) and (\ref{Model-CKT-current}).

In the studies of the second-order responses, one should use the appropriate CKT. In a general case, the kinetic theory including the field-induced corrections to the Berry curvature and quasiparticle dispersion relation was derived in Refs.[S6,S7].
Its explicit formulation in the case of Weyl semimetals is given in Ref.~[S8].
It is notable that the equations of motion, the Boltzmann equation (\ref{Model-CKT-kinetic-equation}), and relations for the charge (\ref{Model-CKT-charge}) and current (\ref{Model-CKT-current}) densities will retain their form. However, the Berry curvature $\mathbf{\Omega}_{\lambda}$, the quasiparticle energy $\epsilon_{\mathbf{p}}$, and the velocity $\mathbf{v}_{\mathbf{p}}$ will be modified. In the case of Weyl fermions, the corresponding expressions read as
\begin{align}
\label{Model-CKT-Omega-second}
&\mathbf{\Omega}_{\lambda} =  \mathbf{\Omega}_{\lambda}^{(0)} +\mathbf{\Omega}_{\lambda}^{(1)}+\dots,
\\
&\epsilon_{\mathbf{p}} = \epsilon_{\mathbf{p}}^{(0)} +\epsilon_{\mathbf{p}}^{(1)}+\epsilon_{\mathbf{p}}^{(2)}+\dots,
\label{Model-CKT-energy-second}
\\
&\mathbf{v}_{\mathbf{p}}=\partial_{\mathbf{p}} \epsilon_{\mathbf{p}}=\mathbf{v}^{(0)}_{\mathbf{p}} +\mathbf{v}^{(1)}_{\mathbf{p}} +\mathbf{v}^{(2)}_{\mathbf{p}} +\dots ,
\label{Model-CKT-v}
\end{align}
where the components are
\begin{eqnarray}
\label{Model-CKT-Omega-second-0}
\mathbf{\Omega}_{\lambda}^{(0)}&=&  \lambda  \eta \hbar \frac{\hat{\mathbf{p}}}{2p^2},\\
\label{Model-CKT-Omega-second-1}
\mathbf{\Omega}_{\lambda}^{(1)}&=& -\frac{e \hbar^2}{4p^4} \left[\frac{2}{c}\hat{\mathbf{p}}(\hat{\mathbf{p}}\cdot\mathbf{B}_{\lambda})
-\frac{1}{c}\mathbf{B}_{\lambda} +\frac{2\eta }{v_F}\left(\tilde{\mathbf{E}}_{\lambda}\times\hat{\mathbf{p}}\right)\right]
\end{eqnarray}
and
\begin{eqnarray}
\label{Model-CKT-energy-second-0}
\epsilon_{\mathbf{p}}^{(0)} &=& \eta v_Fp,\\
\label{Model-CKT-energy-second-1}
\epsilon_{\mathbf{p}}^{(1)} &=& \lambda \frac{e\hbar v_F }{2cp} (\mathbf{B}_{\lambda}\cdot\hat{\mathbf{p}}),\\
\label{Model-CKT-energy-second-2}
\epsilon_{\mathbf{p}}^{(2)} &=& \frac{e^2 \hbar^2}{4cp^3} \left\{\frac{\eta v_F}{4c}
\left[2 B_{\lambda}^2-(\mathbf{B}_{\lambda}\cdot\hat{\mathbf{p}})^2\right] - \left(\mathbf{B}_{\lambda}\cdot\left[\tilde{\mathbf{E}}_{\lambda}\times
\hat{\mathbf{p}}\right]\right)\right\}.
\end{eqnarray}

By making use of Eq.~(\ref{Model-CKT-energy-second}), the quasiparticle velocity $\mathbf{v}^{(n)}_{\mathbf{p}}$ equals
\begin{eqnarray}
\label{Model-CKT-v-0}
\mathbf{v}^{(0)}_{\mathbf{p}} &=& \eta v_F\hat{\mathbf{p}}, \\
\label{Model-CKT-v-1}
\mathbf{v}^{(1)}_{\mathbf{p}} &=& \lambda \frac{ev_F\hbar}{c} \frac{\mathbf{B}_{\lambda}}{2p^2} -\lambda \frac{ev_F\hbar}{c} \hat{\mathbf{p}}\frac{(\hat{\mathbf{p}} \cdot\mathbf{B}_{\lambda})}{p^2},\\
\label{Model-CKT-v-2}
\mathbf{v}^{(2)}_{\mathbf{p}} &=& \frac{5\eta  e^2v_F \hbar^2 \hat{\mathbf{p}} (\mathbf{B}_{\lambda}\cdot\hat{\mathbf{p}})^2}{16c^2p^4}
-\frac{\eta e^2v_F \hbar^2 \mathbf{B}_{\lambda}(\mathbf{B}_{\lambda}\cdot\hat{\mathbf{p}})}{8c^2p^4}
- \frac{3\eta  e^2 \hbar^2v_F}{8c^2p^4}\hat{\mathbf{p}}B_{\lambda}^2 +\frac{e^2 \hbar^2}{4cp^4 }\left(\tilde{\mathbf{E}}_{\lambda}\times
\mathbf{B}_{\lambda}\right)
+\frac{e^2 \hbar^2}{cp^4}\hat{\mathbf{p}} \left(\mathbf{B}_{\lambda}\cdot[\tilde{\mathbf{E}}_{\lambda}\times\hat{\mathbf{p}}]\right). \nonumber\\
\end{eqnarray}
Here $\hat{\mathbf{p}}=\mathbf{p}/p$ and, for simplicity of notations, we omit explicit index $\eta$ at $\epsilon_{\mathbf{p}}$ and $\mathbf{\Omega}_{\lambda}$. Also, we used an effective Hamiltonian of Weyl semimetals in the vicinity of Weyl nodes given in Eq.~(3) in the main text.
It is worth noting that the equations of the CKT presented above include both electromagnetic $\mathbf{E}$ and $\mathbf{B}$ as well as pseudo-electromagnetic $\mathbf{E}_5$ and $\mathbf{B}_5$ fields. As is discussed in the main text, the latter can be induced by strain in Weyl and Dirac semimetals. These fields are expressed through the axial gauge fields $A_{0,5}$ and $\mathbf{A}_5$ as $\mathbf{B}_5=\partial_{\mathbf{r}}\times\mathbf{A}_5$ and $\mathbf{E}_5=-\partial_{\mathbf{r}} A_{0,5}-\partial_t \mathbf{A}_5/c$. As was shown in Refs.~[9,10],
the axial gauge fields are related to the deformation tensor $u_{ij}$ as
\begin{eqnarray}
\label{Model-A5-0}
A_{0,5} &=& -\frac{1}{e} b_0 \beta \sum_{j}u_{jj},\\
\label{Model-A5-vec}
A_{5,i}&=& -\frac{c\hbar b}{e} \Big[\beta u_{iz} +\delta_{iz}\tilde{\beta}(b)\sum_{j}u_{jj}\Big].
\end{eqnarray}
Here $2b_0$ is the separation of Weyl nodes in energy, $b$ is the $z$-component of the momentum-space separation, i.e., the chiral shift (without the loss of generality, we assumed that $\mathbf{b}\parallel \hat{\mathbf{z}}$),
\begin{equation}
\label{Model-transv-uij-def}
u_{ij} = \frac{1}{2} \Big(\partial_ju_i+\partial_iu_j+\sum_{l}\partial_iu_l\partial_ju_l\Big) \approx \frac{1}{2} \left(\partial_ju_i+\partial_iu_j\right)
\end{equation}
is the symmetrized strain tensor, and $\mathbf{u}$ is the displacement vector. The magnitude of strain effects is parameterized by the Gr\"{u}neisen parameter $\beta \equiv -a\partial t/(t\partial a)$, where $t$ is the lattice hopping constant and $a$ is the lattice spacing.
As we will show below, the last term in Eq.~(\ref{Model-A5-vec}) plays an important role when the direction of sound wave propagation and the chiral shift are not aligned. Microscopically, this term is related to the hopping probabilities between the same states (e.g., $s$ or $p$ states in a simple cubic lattice model~[S9,S10]
)
To simplify calculations and present our qualitative results as clear as possible, let us neglect the second order in deformation vector terms in Eq.~(\ref{Model-transv-uij-def}). Such an approximation is also consistent with the linear form of strain-induced axial gauge fields in Eqs.~(\ref{Model-A5-0}) and (\ref{Model-A5-vec}).

The physical {\it consistent} current and charge densities can be represented as a sum of {\it covariant} charge and current as well as the Chern--Simons terms as  $(\rho_{\lambda}, \mathbf{j}_{\lambda}) =\big(\tilde{\rho}_{\lambda}+\rho_{\text{{\tiny CS}}, \lambda}, \tilde{\mathbf{j}}_{\lambda}+\mathbf{j}_{\text{{\tiny CS}}, \lambda}\big)$. The covariant charge and current densities are defined as~~[S2,S4]
\begin{eqnarray}
\label{Model-CKT-charge}
\tilde{\rho}_{\lambda}
&=& -\sum_{\eta=\pm}\eta e\int\frac{d^3p}{(2\pi \hbar)^3} \left[1-\frac{e}{c}(\mathbf{B}_{\lambda}\cdot\mathbf{\Omega}_{\lambda})\right] f_{\lambda},\\
\label{Model-CKT-current}
\tilde{\mathbf{j}}_{\lambda}
&=& -\sum_{\eta=\pm}\eta e\int\frac{d^3p}{(2\pi \hbar)^3}\left\{\mathbf{v}_{\mathbf{p}} -\frac{e}{c}(\mathbf{v}_{\mathbf{p}}\cdot\mathbf{\Omega}_{\lambda}) \mathbf{B}_{\lambda} -e\left(\tilde{\mathbf{E}}_{\lambda}\times\mathbf{\Omega}_{\lambda}\right)\right\} f_{\lambda} \nonumber\\
&&-\sum_{\eta=\pm}\eta e\partial_{\mathbf{r}}\times \int\frac{d^3p}{(2\pi \hbar)^3} f_{\lambda}\epsilon_{\mathbf{p}}\mathbf{\Omega}_{\lambda}.
\end{eqnarray}
Here the last term in Eq.~(\ref{Model-CKT-current}) is the magnetization current and the overall prefactor $\eta$ originates from the fact that the contribution of the filled states (i.e., $\delta_{\eta,-}$ in Eq.~(\ref{Model-CKT-equilibrium-function-eta})) was ignored. The Chern--Simons terms in the electric charge and current densities are~[S11,S12]
\begin{eqnarray}
\label{Model-CKT-consistent-charge-density}
\rho_{\text{{\tiny CS}}} &=&\frac{e^2}{2\pi^2 \hbar c} \left(\mathbf{b}\cdot\mathbf{B}\right) -\frac{e^3}{2\pi^2 \hbar^2c^2} \left(\mathbf{A}_5\cdot\mathbf{B}\right),\\
\label{Model-CKT-consistent-current-density}
\mathbf{j}_{\text{{\tiny CS}}} &=&\frac{e^2}{2\pi^2 \hbar} b_0 \mathbf{B} -\frac{e^3}{2\pi^2 \hbar^2 c} A_{0,5} \mathbf{B} -\frac{e^2}{2\pi^2 \hbar} \left(\mathbf{b}
\times\tilde{\mathbf{E}}\right) +\frac{e^3}{2\pi^2 \hbar^2 c} \left(\mathbf{A}_5
\times\tilde{\mathbf{E}}\right).
\end{eqnarray}
Their analog in the chiral current and charge densities reads as~[S11]
\begin{eqnarray}
\label{Model-CKT-consistent-charge-5-density}
\rho_{\text{{\tiny CS}}, 5} &=& \frac{e^2}{6\pi^2 \hbar c} \left(\mathbf{b}\cdot\mathbf{B}_5\right) -\frac{e^3}{6\pi^2 \hbar^2c^2} \left(\mathbf{A}_5\cdot\mathbf{B}_5\right),\\
\label{Model-CKT-consistent-current-5-density}
\mathbf{j}_{\text{{\tiny CS}}, 5} &=& \frac{e^2}{6\pi^2 \hbar} b_0 \mathbf{B}_5 -\frac{e^3}{6\pi^2 \hbar^2 c} A_{0,5} \mathbf{B}_5 -\frac{e^2}{6\pi^2 \hbar} \left(\mathbf{b} \times\tilde{\mathbf{E}}_5\right) +\frac{e^3}{6\pi^2 \hbar^2 c} \left(\mathbf{A}_5 \times\tilde{\mathbf{E}}_5\right).
\end{eqnarray}
As one of us advocated in Ref.~[S11],
the Chern--Simons terms in the current and charge densities are important to cure the anomalous {\it local} electric charge nonconservation in external electromagnetic and pseudo-electromagnetic fields. The consistent current and charge densities satisfy the following continuity relations
~[S11,S12]:
\begin{align}
\label{Model-CKT-dn/dt-n5}
&\partial_t \rho_5 +\partial_{\mathbf{r}}\cdot \mathbf{j}_5 = -\frac{e^3}{2\pi^2 \hbar^2 c}
\Big[\left(\mathbf{E}\cdot\mathbf{B}\right) +\frac{1}{3}\left(\mathbf{E}_{5}\cdot\mathbf{B}_{5}\right)\Big],\\
\label{Model-CKT-dn/dt-n}
&\partial_t \rho +\partial_{\mathbf{r}}\cdot \mathbf{j} = 0.
\end{align}
Here the nonconservation of chiral charge determined by the right-hand side in Eq.~(\ref{Model-CKT-dn/dt-n5}) is related to the chiral anomaly and its modification by pseudo-electromagnetic fields.

To clarify the possibility of the acoustogalvanic response in Dirac and Weyl semimetals, we consider the case in which the external electromagnetic fields are absent $\mathbf{E}=\mathbf{B}=\mathbf{0}$. Furthermore, we assume that the parity-inversion symmetry is not broken, i.e., $b_0=0$ and two Weyl nodes are separated by $2\mathbf{b}$ in momentum space.
This model setup can be straightforwardly generalized to the case of certain Dirac semimetals, whose low energy spectrum contain Dirac points separated in momentum space. Among them are $A_3$Bi ($A=$ Na, K, Rb) and Cd$_3$As$_2$~[S13,S14],
which can be also considered as $\mathbb{Z}_2$ Weyl semimetals~[S15].
In particular, the second-order contributions from different copies of Weyl semimetals should simply add up in these systems.

We assume a longitudinal propagation of a sound wave with the displacement vector $\mathbf{u} = \frac{1}{2} u_{0} \hat{\mathbf{q}} e^{-i\omega t +i\mathbf{q}\mathbf{r}} +c.c.$, where $\omega$ is the sound frequency, $q=\omega/v_s$ is the absolute value of the wave vector, and $v_s$ is the sound velocity.
Then, by using the gauge potential in Eq.~(\ref{Model-A5-vec}), the pseudoelectric and pseudomagnetic fields are
\begin{eqnarray}
\label{App-Model-E5-def}
E_{5,j} = -\frac{i\hbar\omega b}{2e} \Big[\beta u_{jz} +\delta_{jz}\tilde{\beta}(b)\sum_{l}u_{ll}\Big] +c.c. 
= \frac{\hbar\omega b u_0}{2e} \Big[\beta \frac{q_j q_z}{q} +\delta_{jz}\tilde{\beta}(b)q\Big]  e^{-i\omega t +i\mathbf{q}\mathbf{r}}+c.c.,
\end{eqnarray}
and
\begin{eqnarray}
\label{App-Model-B5-def}
B_{5,j} &=& -\frac{c\hbar}{2e} i\epsilon_{jmn}q_m b\Big[\beta u_{nz} +\delta_{nz}\tilde{\beta}(b)\sum_{l}u_{ll}\Big] +c.c. =
\frac{c\hbar}{2e} \epsilon_{jmn}q_m b\Big[\beta \frac{q_n q_z}{q} +\delta_{nz}\tilde{\beta}(b)q\Big]  e^{-i\omega t +i\mathbf{q}\mathbf{r}}+c.c. \nonumber\\
&=& \frac{c\hbar u_0}{2e} \epsilon_{jmz}q_m b\tilde{\beta}(b)q e^{-i\omega t +i\mathbf{q}\mathbf{r}}+c.c.
\end{eqnarray}
For the sake of brevity, we denote these fields as
$\mathbf{E}_5 = \frac{1}{2}\mathbf{E}_{5,0} e^{-i\omega t +i\mathbf{q}\mathbf{r}} +c.c.$ and $\mathbf{B}_5 = \frac{1}{2} \mathbf{B}_{5,0} e^{-i\omega t +i\mathbf{q}\mathbf{r}} +c.c.$, where the subscript $0$ denotes the amplitude of oscillating fields. It is worth noting that, as follows from Eqs.~(\ref{App-Model-E5-def}) and (\ref{App-Model-B5-def}), pseudoelectric and pseudomagnetic fields are orthogonal, $\left(\mathbf{E}_{5}\cdot\mathbf{B}_{5}\right)=0$.

Like in conventional materials, deformations in Weyl and Dirac semimetals lead not only to the pseudo-electromagnetic fields $\mathbf{E}_5$ and $\mathbf{B}_5$, but modify the quasiparticle energy $\epsilon_{\mathbf{p}}$ due to the deformation potential term $\sum_{i,j}D_{ij}=\sum_{n=1}\sum_{i,j}D_{ij}^{(n)}(\mathbf{p})$~[S16--S18],
i.e.,
\begin{eqnarray}
\label{app-cc-dp-eps}
\epsilon_{\mathbf{p}} \to \epsilon_{\mathbf{p}} + \sum_{n=1}\sum_{i,j}D_{ij}^{(n)}(\mathbf{p}).
\end{eqnarray}
In the first order $n=1$, $D_{ij}^{(1)}(\mathbf{p})\propto u_{ij}$.
In order to simplify our calculations, we assume that the deformation potential does not depend on momentum, $D_{ij}(\mathbf{p}) \approx D_{ij}$. Further, for our qualitative estimates, it is sufficient to ignore anisotropy, i.e., $D_{ij}\approx D \delta_{ij}$. As will be shown below, the conventional acoustoelectric effect will be absent in this case.

The propagation of sound waves distorts the ionic lattice leading to deviations of local electric charge density from its equilibrium value. These deviations are (partially) screened by free charge carriers, which is captured by the following corrections to the chemical potential:
\begin{eqnarray}
\label{app-cc-dp-SK-screen-mu}
\mu_{\lambda} \to
\mu_{\lambda} = \sum_{n=0}\mu_{\lambda}^{(n)},
\end{eqnarray}
where $\mu_{\lambda}^{(0)} = \mu$.
In general, deviations of the chiral chemical potential are also allowed in chiral systems such as Weyl semimetals.
Up to the second order in weak perturbations, we have the following {\it local equilibrium} distribution function:
\begin{eqnarray}
\label{app-cc-dp-SK-screen-f}
f^{(0)}_{\lambda}
&=& \frac{1}{1+e^{\eta \left(\epsilon_{\mathbf{p}} -\mu_{\lambda} +V_{\lambda}\right)/T}} \approx f^{\rm eq}_{\lambda} + \Big(\epsilon_{\mathbf{p}}^{(1)}+V_{\lambda}^{(1)}\Big) \partial_{\epsilon_{\mathbf{p}}}f^{\rm eq}_{\lambda} +
\frac{\Big(\epsilon_{\mathbf{p}}^{(1)}+V_{\lambda}^{(1)}\Big)^2}{2} \partial_{\epsilon_{\mathbf{p}}}^2f^{\rm eq}_{\lambda} + \Big(\epsilon_{\mathbf{p}}^{(2)}+V_{\lambda}^{(2)}\Big)\partial_{\epsilon_{\mathbf{p}}}f^{\rm eq}_{\lambda},
\end{eqnarray}
where $V_{\lambda}^{(n)} = D^{(n)}  - \mu_{\lambda}^{(n)}$
and the global equilibrium distribution function $f^{\rm eq}_{\lambda}$ is given in Eq.~(\ref{Model-CKT-equilibrium-function}).

\section{S II. Current density}
\label{sec:charge-current}

In this section, the current density up to the second order in strain-induced pseudo-electromagnetic fields is calculated. For our estimates, we use the numerical parameters valid for the Dirac semimetal Cd$_3$As$_2$. They are~
[S19-S23]
\begin{equation}
\label{app-model-num-par}
v_{\rm F} \approx 1.5\times 10^{8}~\mbox{cm/s}, \quad b \approx 1.6~\mbox{nm}^{-1}, \quad \mu_0\approx 200~\mbox{meV}, \quad \tau_0= 1~\mbox{ps}, \quad v_s\approx 2.3\times 10^{5}~\mbox{cm/s}.
\end{equation}
In addition, we assume that $\mu_5=0$, $T=5~\mbox{K}$, $\beta\approx1$, and $\tilde{\beta}\approx 1$. It is worth noting that since the single-band approximation is used, the electric chemical potential should be significantly higher than the frequency of sound, i.e., $\mu\gg \hbar \omega$, which is indeed the case for many realistic Weyl and Dirac semimetals. By using the numerical parameters in Eq.~(\ref{app-model-num-par}), we estimate that $\mu/\hbar\approx 48.36~\mbox{THz}$, which is well above typical ultrasound frequencies.

\subsection{A. First order response}
\label{sec:charge-current-first}

In the {\it first order} in weak fields, we have the following Boltzmann equation:
\begin{eqnarray}
&&\partial_t f_{\lambda}^{(1)} -e\Big[\tilde{\mathbf{E}}_{\lambda} +\frac{1}{c}\left(\mathbf{v}_{\mathbf{p}}^{(0)}\times \mathbf{B}_{\lambda}\right)\Big]\cdot
\partial_\mathbf{p} f_{\lambda}^{(1)} +\left(\mathbf{v}_{\mathbf{p}}^{(0)}\cdot \partial_\mathbf{r} f_{\lambda}^{(1)}\right) 
+(\partial_{\epsilon_{\mathbf{p}}}f_{\lambda}^{\rm eq}) \partial_t \left(\epsilon_{\mathbf{p}}^{(1)}+V_{\lambda}^{(1)}\right) = -\frac{f_{\lambda}^{(1)}}{\tau}.
\label{CKT-First-eq}
\end{eqnarray}
It is straightforward to find the following solution for the above equation:
\begin{eqnarray}
\label{CKT-First-sol-def}
f_{\lambda}^{(1)}&=& \frac{1}{2} f_{\lambda,0} e^{-i\omega t +i\mathbf{q}\mathbf{r}} +c.c.,
\end{eqnarray}
where the amplitude of the distribution function reads as
\begin{eqnarray}
\label{CKT-First-sol}
f_{\lambda,0}^{(1)}&=& \frac{1}{2} \frac{e\tau\left(\tilde{\mathbf{E}}_{\lambda,0} +\frac{1}{c}\left[\mathbf{v}_{\mathbf{p}}^{(0)}\times \mathbf{B}_{\lambda,0}\right)\right] (\partial_{\mathbf{p}} f_{\lambda}^{\rm eq}) +i\omega \tau \left(V_{\lambda,0}^{(1)}+\epsilon_{\mathbf{p},0}^{(1)}\right) (\partial_{\epsilon_{\mathbf{p}}}f_{\lambda}^{\rm eq})}{1-i\omega \tau +i(\mathbf{v}_{\mathbf{p}}^{(0)}\cdot\mathbf{q})\tau} \nonumber\\
&\approx& \frac{1}{2} \left\{e\tau\left[\tilde{\mathbf{E}}_{\lambda,0} +\frac{1}{c}\left(\mathbf{v}_{\mathbf{p}}^{(0)}\times \mathbf{B}_{\lambda,0}\right)\right](\partial_{\mathbf{p}} f_{\lambda}^{\rm eq}) +i\omega \tau \left(V_{\lambda,0}^{(1)}+\epsilon_{\mathbf{p},0}^{(1)}\right) (\partial_{\epsilon_{\mathbf{p}}}f_{\lambda}^{\rm eq}) \right\} \left[1+i\omega \tau -i(\mathbf{v}_{\mathbf{p}}^{(0)}\cdot\mathbf{q})\tau \left(1+2\omega\tau\right)\right]. \nonumber\\
\end{eqnarray}
In the last expression, for simplicity, we assumed that both $\omega \tau \ll1$ and $v_F \omega \tau/v_s \ll1$ are small and expanded the denominator. According to the numerical parameters given in Eq.~(\ref{app-model-num-par}), this approximation is indeed reasonable for realistic values of $\omega$ and $\tau$.

The charge and current densities in the first order in the fields read as
\begin{eqnarray}
\label{CKT-First-rho}
\rho_{\lambda}^{(1)} &=& -e\sum_{\eta=\pm} \eta \int\frac{d^3p}{(2\pi\hbar)^3} f_{\lambda}^{(1)} - e\sum_{\eta=\pm} \eta\int\frac{d^3p}{(2\pi\hbar)^3} \left[-\frac{e}{c} \left(\mathbf{B}_{\lambda}\cdot\bm{\Omega}_{\lambda}^{(0)}\right) f_{\lambda}^{\rm eq} + \left(V_{\lambda}^{(1)}+ \epsilon_{\mathbf{p}}^{(1)}\right) (\partial_{\epsilon_{\mathbf{p}}}f_{\lambda}^{\rm eq})\right]
\end{eqnarray}
and
\begin{eqnarray}
\label{CKT-First-J}
j_{\lambda}^{(1)} &=& -e\sum_{\eta=\pm} \eta\int\frac{d^3p}{(2\pi\hbar)^3} \mathbf{v}_{\mathbf{p}}^{(0)}f_{\lambda}^{(1)} - e\sum_{\eta=\pm} \eta\int\frac{d^3p}{(2\pi\hbar)^3} i \left(\mathbf{q}\times \bm{\Omega}_{\lambda}^{(0)}\right) \epsilon_{\mathbf{p}}^{(0)} f_{\lambda}^{(1)} \nonumber\\
&-& e\sum_{\eta=\pm} \eta\int\frac{d^3p}{(2\pi\hbar)^3} \left[\mathbf{v}_{\mathbf{p}}^{(1)} -\frac{e}{c} \mathbf{B}_{\lambda}\left(\mathbf{v}_{\mathbf{p}}^{(0)}\cdot \bm{\Omega}_{\lambda}^{(0)}\right) - e\left(\tilde{\mathbf{E}}_{\lambda}\times \bm{\Omega}_{\lambda}^{(0)}\right)\right] \left[f_{\lambda}^{\rm eq} +\left(V_{\lambda}^{(1)}+ \epsilon_{\mathbf{p}}^{(1)}\right) (\partial_{\epsilon_{\mathbf{p}}}f_{\lambda}^{\rm eq})\right],
\end{eqnarray}
respectively. By using expressions in Sec.~\ref{sec:App-formulas} for calculating the integrals over angles and momenta, it is straightforward to obtain the following amplitudes of the oscillating charge and current densities:
\begin{eqnarray}
\label{CKT-First-rho-answer}
\rho_{\lambda,0}^{(1)} &=& -i\frac{e^2 \tau^2 v_F^2\left(\mathbf{E}_{\lambda}\cdot\mathbf{q}\right)}{3} \left(1+2i\omega\tau\right) C_1 + e V_{\lambda}^{(1)} \left(1+i\omega\tau -\omega^2 \tau^2\right) C_1
\end{eqnarray}
and
\begin{eqnarray}
\label{CKT-First-j-answer}
\mathbf{j}_{\lambda,0}^{(1)} &=& \frac{e^2 \tau v_F \mathbf{E}_{\lambda}}{3} \left(1+i\omega\tau\right) C_1 + \frac{e\omega\tau^2 v_F^2 V_{\lambda}^{(1)} \mathbf{q}}{3} \left(1+2i\omega\tau\right) C_1 + i\lambda\frac{e^2 \hbar \omega \tau v_F^2 \mathbf{B}_{\lambda}}{6c} \left(1+i\omega\tau\right)C_2 \nonumber\\
&-&i \lambda \frac{e^2 \hbar \tau v_F^2 \left(\mathbf{q}\times \mathbf{E}_{\lambda}\right)}{6} \left(1+i\omega\tau\right) C_2
-\frac{e^2\hbar^2v_F^3 \omega \tau \left(\mathbf{q}\times \mathbf{B}_{\lambda}\right)}{12c} C_3
-\lambda\frac{e^2\hbar v_F^2 \mathbf{B}_{\lambda}}{2c} C_2,
\end{eqnarray}
respectively. The shorthand notations are
\begin{eqnarray}
\label{CKT-First-C1}
C_1 &=& \sum_{\eta=\pm} \int\frac{d^3p}{(2\pi \hbar)^3} \partial_{\epsilon_{\mathbf{p}}} f^{\rm eq}_{\lambda}= -\frac{2}{v_F}\sum_{\eta=\pm} \eta\int\frac{d^3p}{(2\pi \hbar)^3} \frac{1}{p} f^{\rm eq}_{\lambda} = \frac{1}{2\pi^2 \hbar^3 v_F^3} \left(\mu_{\lambda}^2 + \frac{\pi^2 T^2}{3}\right),\\
\label{CKT-First-C2}
C_2 &=& \sum_{\eta=\pm} \eta \int\frac{d^3p}{(2\pi \hbar)^3} \frac{1}{p}\partial_{\epsilon_{\mathbf{p}}} f^{\rm eq}_{\lambda}= -\frac{1}{v_F}\sum_{\eta=\pm} \int\frac{d^3p}{(2\pi \hbar)^3} \frac{1}{p^2} f^{\rm eq}_{\lambda} = -\frac{\mu_{\lambda}}{2\pi^2 \hbar^3 v_F^2},\\
\label{CKT-First-C3}
C_3 &=& \sum_{\eta=\pm} \eta \int\frac{d^3p}{(2\pi \hbar)^3} \frac{1}{p^2}\partial_{\epsilon_{\mathbf{p}}} f^{\rm eq}_{\lambda} = -\frac{1}{v_F}\sum_{\eta=\pm}  \int\frac{d^3p}{(2\pi \hbar)^3} \frac{1}{p^3}f^{\rm eq}_{\lambda} =-v_F\sum_{\eta=\pm} \int\frac{d^3p}{(2\pi \hbar)^3} \frac{1}{p}\partial_{\epsilon_{\mathbf{p}}}^2 f^{\rm eq}_{\lambda} = -\frac{1}{2\pi^2 \hbar^3 v_F}.
\end{eqnarray}

To determine the correction to the chemical potential $\mu^{(1)}_{\lambda}$ (recall that $V_{\lambda}^{(i)} = D^{(i)}-\mu^{(i)}_{\lambda}$), we enforce the continuity relation for both electric and chiral currents,
\begin{eqnarray}
\label{CKT-First-cont-rel}
\partial_t \rho^{(1)} + \partial_{\mathbf{r}}\cdot\mathbf{j}^{(1)}  = 0,\\
\label{CKT-First-cont-rel-5}
\partial_t \rho_5^{(1)} + \partial_{\mathbf{r}}\cdot\mathbf{j}_5^{(1)}  = 0.
\end{eqnarray}
Note that since $ \mathbf{E}_5\cdot\mathbf{B}_5 =0$, the chiral charge is also conserved.
It is straightforward to check that the first equation leads to $V^{(1)}=0$. This means that the deformation potential $D^{(1)}$ is completely compensated by the deviations of the electric chemical potential $\mu^{(1)}$ in the model at hand. On the other hand, Eq.~(\ref{CKT-First-cont-rel-5}) allows for the following nontrivial solution:
\begin{eqnarray}
\label{CKT-First-V-5}
V_{5,0}^{(1)}= -\mu_{5,0}^{(1)} = \frac{e\tau v_F^2}{3\omega} \left(\mathbf{E}_5\cdot\mathbf{q}\right) \frac{1-2\omega^2\tau^2 +2i\omega\tau}{1-\omega^2\tau^2 +i\omega\tau -\frac{v_F^2q^2\tau^2}{3} (1+2i\omega\tau)} \approx \frac{e\tau v_F^2}{3\omega} \left(\mathbf{E}_5\cdot\mathbf{q}\right) \left(1+i\omega\tau\right)+O(\omega^2\tau^2).
\end{eqnarray}

In the case of an exact expression for the distribution function given in the first line in Eq.~(\ref{CKT-First-sol}) and at $\mathbf{q}\parallel\hat{\mathbf{z}}$, the amplitude $V_{5,0}^{(1)}$ reads as
\begin{eqnarray}
\label{CKT-First-V-5-exact}
V_{5,0}^{(1)} = -\frac{v_F\tau eE_{5,0} \left[\omega K_1(A_1,A_2) -v_Fq_z K_2(A_1,A_2)\right]}{2\omega+i\omega \tau  \left[\omega K_0(A_1,A_2) -v_F q_z K_1(A_1,A_2) \right]}.
\end{eqnarray}
Here
\begin{eqnarray}
\label{CKT-First-angle-integral-K0}
K_0(A_1,A_2) &=& \int_{-1}^{1} d\cos{\theta} \frac{1}{A_1+iA_2 \cos{\theta}} = \frac{2}{A_2} \arctan{\left(\frac{A_2}{A_1}\right)},\\
\label{CKT-First-angle-integral-K1}
Q_1(A_1,A_2) &=& \int_{-1}^{1} d\cos{\theta} \frac{\cos{\theta}}{A_1+iA_2 \cos{\theta}} = -\frac{2i}{A_2^2} \left[A_2 - A_1 \arctan{\left(\frac{A_2}{A_1}\right)}\right],\\
\label{CKT-First-angle-integral-K2}
K_2(A_1,A_2) &=& \int_{-1}^{1} d\cos{\theta} \frac{\cos^2{\theta}}{A_1+iA_2 \cos{\theta}} =
\frac{2A_1}{A_2^3} \left[A_2 - A_1 \arctan{\left(\frac{A_2}{A_1}\right)}\right],
\end{eqnarray}
and we used $A_1 = 1-i\omega \tau$ and $A_2 = v_F q_z\tau$. Expression (\ref{CKT-First-V-5-exact}) is valid even for $\omega \tau \approx1$.

By using Eq.~(\ref{CKT-First-j-answer}), we derive the following amplitude of the oscillating electric current in the first order in deformations:
\begin{eqnarray}
\label{CKT-First-j-electric}
\mathbf{j}_{0}^{(1)} &=& \sum_{\lambda=\pm}\mathbf{j}_{\lambda,0}^{(1)} \approx
\frac{e^2 \mu \mathbf{B}_{5}}{6\pi^2 \hbar^2c} \left(1-i\omega \tau\right)
+i \frac{e^2 \tau \mu \left(\mathbf{q}\times \mathbf{E}_{5}\right)}{6\pi^2 \hbar^2} \left(1+i\omega\tau\right) + {\cal O}[(\omega\tau)^2].
\end{eqnarray}

In addition to the electric current density, dynamic pseudo-electromagnetic fields lead to the chiral current
\begin{eqnarray}
\label{CKT-First-j5-electric}
\mathbf{j}_{5,0}^{(1)} &=& \sum_{\lambda=\pm}\lambda \mathbf{j}_{\lambda,0}^{(1)} \approx
\frac{2e^2\tau \mathbf{E}_5}{6\pi^2 \hbar^3 v_F^2}\left(1+i\omega \tau\right) \left(\mu^2 + \frac{\pi^2 T^2}{3}\right) +\frac{e^2 v_F \tau^3 \mathbf{q} \left(\mathbf{q}\cdot\mathbf{E}_5\right)}{9\pi^2 \hbar^3}\left(1+3i\omega \tau\right) \left(\mu^2 + \frac{\pi^2 T^2}{3}\right) \nonumber\\
&+& \frac{e^2 v_F^2 \omega \tau\left(\mathbf{q}\times \mathbf{B}_5\right)}{12\pi^2 \hbar c} + {\cal O}[(\omega\tau)^2].
\end{eqnarray}

\subsection{B. Second order response}
\label{sec:charge-current-second}

Let us consider the {\it second order} response to the pseudo-electromagnetic fields $\mathbf{E}_5$ and $\mathbf{B}_5$. Note that since we used the strain tensor and the axial gauge fields in the linear approximation, the results for the second-order acoustogalvanic response should be considered as qualitative rather than quantitative.
The second-order Boltzmann equation reads as
\begin{eqnarray}
\label{CKT-Second-Boltzmann-Eq}
&&-\frac{e}{c}\left(\mathbf{B}_{\lambda}\cdot\bm{\Omega}_{\lambda}^{(0)}\right) \partial_t\left[f_{\lambda}^{(1)} +\left(V_{\lambda}^{(1)} +\epsilon_{\mathbf{p}}^{(1)}\right) \partial_{\epsilon_{\mathbf{p}}}f_{\lambda}^{\rm eq}\right]
-e\left[\tilde{\mathbf{E}}_{\lambda} +\frac{1}{c}\left(\mathbf{v}_{\mathbf{p}}^{(0)}\times \mathbf{B}_{\lambda}\right)\right] \partial_{\mathbf{p}}\left[f_{\lambda}^{(1)} +\left(V_{\lambda}^{(1)} +\epsilon_{\mathbf{p}}^{(1)}\right) \partial_{\epsilon_{\mathbf{p}}}f_{\lambda}^{\rm eq}\right] \nonumber\\
&&-\frac{e}{c} \left(\mathbf{v}_{\mathbf{p}}^{(1)}\times\mathbf{B}_{\lambda}\right) \partial_{\mathbf{p}} f_{\lambda}^{\rm eq}
+\left[\mathbf{v}_{\mathbf{p}}^{(1)}-e\left(\tilde{\mathbf{E}}\times \bm{\Omega}_{\lambda}^{(0)}\right) -\frac{e}{c}\left(\mathbf{v}_{\mathbf{p}}^{(0)}\cdot\bm{\Omega}_{\lambda}^{(0)}\right)\right] \partial_{\mathbf{r}} f_{\lambda}^{(1)}
= -\frac{f^{(2)}}{\tau}
\end{eqnarray}
By using the definition of $\mathbf{v}_{\mathbf{p}}^{(1)}$ in Eq.~(\ref{Model-CKT-v-1}), it is straightforward to show that the term $\big(\mathbf{v}_{\mathbf{p}}^{(1)}\times\mathbf{B}_{\lambda}\big) \partial_{\mathbf{p}} f_{\lambda}^{\rm eq}$ in the second line in Eq.~(\ref{CKT-Second-Boltzmann-Eq}) vanishes. Then, by solving Eq.~(\ref{CKT-Second-Boltzmann-Eq}), we obtain the following amplitude of the distribution function that corresponds to the direct current (dc) response:
\begin{eqnarray}
\label{CKT-Second-sol}
f_{\lambda,0}^{(2)} &=& -\frac{\tau}{4} \Bigg\{ -e\left[\tilde{\mathbf{E}}_{\lambda,0} +\frac{1}{c} \left(\mathbf{v}_{\mathbf{p}}^{(0)}\times \mathbf{B}_{\lambda,0}\right)\right]^{*} (\partial_{\mathbf{p}} f_{\lambda,0}^{(1)}) +i\mathbf{q}\cdot\left[\mathbf{v}_{\mathbf{p}}^{(1)} -e \left(\tilde{\mathbf{E}}_{\lambda,0}\times \bm{\Omega}_{\lambda}^{(0)}\right) -\frac{e}{c}\left(\mathbf{v}_{\mathbf{p}}^{(0)} \cdot\bm{\Omega}_{\lambda}^{(0)}\right) \mathbf{B}_{\lambda,0}\right]^{*} f_{\lambda,0}^{(1)}
\nonumber\\
&+&\frac{ie\omega}{c} \left(\mathbf{B}_{\lambda,0}^{*}\cdot\bm{\Omega}_{\lambda}^{(0)}\right) f_{\lambda,0}^{(1)}
-e\left[\tilde{\mathbf{E}}_{\lambda,0} + \frac{1}{c}\left(\mathbf{v}_{\mathbf{p}}^{(0)}\times \mathbf{B}_{\lambda,0}\right) \right]^{*} (\partial_{\epsilon_{\mathbf{p}}} f_{\lambda}^{\rm eq}) \partial_{\mathbf{p}} \left(V_{\lambda,0}^{(1)} +\epsilon_{\mathbf{p},0}^{(1)}\right) \nonumber\\
&+&\frac{ie\omega}{c} \left(\mathbf{B}_{\lambda,0}\cdot\bm{\Omega}_{\lambda}^{(0)}\right)^{*} \left(V_{\lambda,0}^{(1)} +\epsilon_{\mathbf{p},0}^{(1)}\right) (\partial_{\epsilon_{\mathbf{p}}} f_{\lambda}^{\rm eq})
\Bigg\} +c.c.
\end{eqnarray}
Here the derivative with respect to momentum from $f_{\lambda,0}^{(1)}$ is
\begin{eqnarray}
\label{CKT-Second-f1-part-p}
\partial_{\mathbf{p}} f_{\lambda,0}^{(1)} &=& e\eta \tau v_F \left[1+i\omega\tau -iv_F \tau \left(\mathbf{q}\cdot\hat{\mathbf{p}}\right) \left(1+2i\omega \tau\right)\right]
\Bigg\{\left[\mathbf{E}_{\lambda,0} -\hat{\mathbf{p}}\left(\mathbf{E}_{\lambda,0}\cdot\hat{\mathbf{p}}\right)\right] \frac{(\partial_{\epsilon_{\mathbf{p}}} f_{\lambda}^{\rm eq})}{p} + \eta v_F \hat{\mathbf{p}} \left(\mathbf{E}_{\lambda,0}\cdot\hat{\mathbf{p}}\right) (\partial_{\epsilon_{\mathbf{p}}}^2 f_{\lambda}^{\rm eq})
\Bigg\} \nonumber\\
&-& ie\eta \tau^2 v_F^2 \left(\mathbf{E}_{\lambda,0}\cdot \hat{\mathbf{p}}\right) \left[\mathbf{q} - \hat{\mathbf{p}}\left(\mathbf{q}\cdot\hat{\mathbf{p}}\right)\right] \left(1+2i\omega \tau\right) \frac{(\partial_{\epsilon_{\mathbf{p}}} f_{\lambda}^{\rm eq})}{p}
+i\lambda \eta \frac{e\hbar v_F \omega \tau}{2c} \left[1+i\omega\tau -iv_F \tau \left(\mathbf{q}\cdot\hat{\mathbf{p}}\right) \left(1+2i\omega \tau\right)\right] \nonumber\\ &\times&\left[\mathbf{B}_{\lambda,0} -2\hat{\mathbf{p}}\left(\hat{\mathbf{p}}\cdot\mathbf{B}_{\lambda,0}\right) \right] \frac{(\partial_{\epsilon_{\mathbf{p}}} f_{\lambda}^{\rm eq})}{p^2} + \lambda \eta \frac{\omega \tau^2 v_F^2}{2c} \left(\mathbf{B}_{\lambda,0}\cdot\hat{\mathbf{p}}\right) \left[\mathbf{q} - \hat{\mathbf{p}}\left(\mathbf{q}\cdot\hat{\mathbf{p}}\right)\right] \left(1+2i\omega \tau\right) \frac{(\partial_{\epsilon_{\mathbf{p}}} f_{\lambda}^{\rm eq})}{p^2} \nonumber\\
&+&i\lambda \frac{e\hbar \omega \tau v_F^2}{2c} \hat{\mathbf{p}} \left(\mathbf{B}_{\lambda,0}\cdot\hat{\mathbf{p}}\right) \left[1+i\omega\tau -iv_F \tau \left(\mathbf{q}\cdot\hat{\mathbf{p}}\right) \left(1+2i\omega \tau\right)\right] \frac{(\partial_{\epsilon_{\mathbf{p}}}^2 f_{\lambda}^{\rm eq})}{p} \nonumber\\
&+&i\eta\omega \tau v_F V_{\lambda,0}^{(1)} \hat{\mathbf{p}} (\partial_{\epsilon_{\mathbf{p}}}^2 f_{\lambda}^{\rm eq}) \left[1+i\omega\tau -iv_F \tau \left(\mathbf{q}\cdot\hat{\mathbf{p}}\right) \left(1+2i\omega \tau\right)\right] +v_F \omega \tau^2 V_{\lambda,0}^{(1)} \frac{(\partial_{\epsilon_{\mathbf{p}}} f_{\lambda}^{\rm eq})}{p}  \left[\mathbf{q} - \hat{\mathbf{p}}\left(\mathbf{q}\cdot\hat{\mathbf{p}}\right)\right] \left(1+2i\omega \tau\right). \nonumber\\
\end{eqnarray}
Since we consider the dc response, continuity relations for electric and chiral current densities are automatically fulfilled. Therefore, the second order correction to the chemical potential $\mu^{(2)}$ can be determined from the condition of the electric charge neutrality $\rho^{(2)}=0$. The corresponding correction does not provide any contributions to the dc current and, therefore, will not be considered.

The general expression for the second-order rectified current reads as
\begin{eqnarray}
\label{CKT-Second-j-def}
j_{\lambda}^{(2)} &=& -e\sum_{\eta=\pm} \frac{\eta}{4} \int \frac{d^3p}{(2\pi \hbar)^3} \Bigg\{
2\mathbf{v}_{\mathbf{p}}^{(0)} f_{\lambda}^{(2)} +\left(\mathbf{v}_{\mathbf{p},0}^{(1)}\right)^{*} f_{\lambda,0}^{(1)}
-\frac{e}{c}\left(\mathbf{v}_{\mathbf{p}}^{(0)}\cdot\bm{\Omega}_{\lambda}^{(0)}\right) \mathbf{B}_{\lambda,0}^{*} f_{\lambda,0}^{(1)}
-e\left(\tilde{\mathbf{E}}_{\lambda,0}\times \bm{\Omega}_{\lambda}^{(0)}\right)^{*} f_{\lambda,0}^{(1)} \nonumber\\
&+&v_{\mathbf{p}}^{(2)} f_{\lambda}^{\rm eq}
-\frac{e}{c} \left(\mathbf{v}_{\mathbf{p},0}^{(1)}\cdot\bm{\Omega}_{\lambda}^{(0)}\right) \mathbf{B}_{\lambda,0}^{*} f_{\lambda}^{\rm eq}
-\frac{e}{c} \left(\mathbf{v}_{\mathbf{p}}^{(0)}\cdot\bm{\Omega}_{\lambda,0}^{(1)}\right) \mathbf{B}_{\lambda,0}^{*} f_{\lambda}^{\rm eq}
-e\left(\mathbf{E}_{\lambda}^{(2)}\times \bm{\Omega}_{\lambda}^{(0)}\right) f_{\lambda}^{\rm eq}
-e\left[\tilde{\mathbf{E}}_{\lambda,0}\times\left(\bm{\Omega}_{\lambda,0}^{(1)}\right)^{*} \right] f_{\lambda}^{\rm eq} \nonumber\\
&+&\left(\mathbf{v}_{\mathbf{p},0}^{(1)}\right)^{*} \left(V_{\lambda,0}^{(1)} +\epsilon_{\mathbf{p},0}^{(1)}\right) f_{\lambda}^{\rm eq}
-\frac{e}{c} \left(\mathbf{v}_{\mathbf{p}}^{(0)}\cdot\bm{\Omega}_{\lambda}^{(0)}\right) \mathbf{B}_{\lambda,0}^{*} \left(V_{\lambda,0}^{(1)} +\epsilon_{\mathbf{p},0}^{(1)}\right) (\partial_{\epsilon_{\mathbf{p}}}f_{\lambda}^{\rm eq}) \nonumber\\
&-& e\left(\tilde{\mathbf{E}}_{\lambda,0}\times\bm{\Omega}_{\lambda}^{(0)}\right) \left(V_{\lambda,0}^{(1)} +\epsilon_{\mathbf{p},0}^{(1)}\right) (\partial_{\epsilon_{\mathbf{p}}}f_{\lambda}^{\rm eq}) +\mathbf{v}_{\mathbf{p}}^{(0)} \frac{1}{2} \left(V_{\lambda,0}^{(1)} +\epsilon_{\mathbf{p},0}^{(1)}\right)^{*}\left(V_{\lambda,0}^{(1)} +\epsilon_{\mathbf{p},0}^{(1)}\right) (\partial_{\epsilon_{\mathbf{p}}}^2f_{\lambda}^{\rm eq})
+\mathbf{v}_{\mathbf{p}}^{(0)} \epsilon_{\mathbf{p}}^{(2)} (\partial_{\epsilon_{\mathbf{p}}}f_{\lambda}^{\rm eq})
\Bigg\} +c.c. \nonumber\\
\end{eqnarray}
After straightforward but tedious calculation, we derive the following current density $\mathbf{j}^{\rm dc} = \sum_{\lambda=\pm} \mathbf{j}_{\lambda}^{(2)}$:
\begin{equation}
\label{CKT-Second-j-DC-1}
j^{\rm dc}_a = \sigma_{abc} E_{5,b} E^{*}_{5,c} +\kappa_{abc} \frac{1}{2}\left(E_{5,b} B^{*}_{5,c}+E_{5,b}^{*} B_{5,c}\right) +\gamma_{abc} B_{5,b} B^{*}_{5,c}.
\end{equation}
Here the response tensors read as
\begin{eqnarray}
\label{CKT-Second-sigma-tens-0}
\sigma_{abc} &=& -\frac{e}{4}\sum_{\lambda=\pm} \Bigg\{
-q_a \delta_{bc} \frac{ie^2v_F^3 \tau^3}{30} \left(1+2i\omega\tau\right) C_2
-\delta_{ac}q_b \frac{e\tau v_F^2}{3}\left[ \frac{4ev_F \tau^3 \omega}{5} - i\omega \tau \tilde{V}_{\lambda}^{(1)} \left(1+i\omega\tau\right)
+\tilde{V}_{\lambda}^{(1)}
\right] C_2 \nonumber\\
&+&q_a q_b q_c \frac{\tau v_F^2}{3} \tilde{V}_{\lambda}^{(1)}\left[ \frac{2ev_F \tau^2}{5} \left(1+2i\omega\tau\right) + \omega \tau \tilde{V}_{\lambda}^{(1)} \left(1+i\omega\tau\right) +i\tilde{V}_{\lambda}^{(1)} \right] C_2 \nonumber\\
&+&\delta_{ab}q_c q^2 \left(\tilde{V}_{\lambda}^{(1)}\right)^{*} \frac{2ev_F^3 \tau^3}{15} \left(1+2i\omega\tau\right) C_2
+\epsilon_{ajb}q_j q_c \left(\tilde{V}_{\lambda}^{(1)}\right)^{*} \frac{i\lambda ev_F \hbar \tau}{6} \left[ 1-2(\omega \tau)^2 \right] C_3
\Bigg\} +c.c.,
\end{eqnarray}
\begin{eqnarray}
\label{CKT-Second-kappa-tens-0}
\kappa_{abc} &=& -\frac{e}{4}\sum_{\lambda=\pm} \Bigg\{
\delta_{ac}q_b \frac{\lambda e\hbar v_F}{30c}
\Big[ i\frac{v_F^2 \tau^2}{2} \left(18+31i \omega\tau -2\omega^2\tau^2\right)
-\tilde{V}_{\lambda}^{(1)} \left(5(3+2\omega^2\tau^2) -5i\omega\tau(1-\omega^2\tau^2) -2v_F^4\omega^4\tau^4 q^2\right)
\Big]\nonumber\\
&+&\epsilon_{abc} \frac{e^2}{c} \left[-\frac{v_F^3\tau^2}{6} \left(1+i\omega\tau\right) C_2 +\frac{i\omega \tau v_F \hbar^2}{12} \left(1-i\omega \tau\right) F +\frac{e^2\hbar^2}{2}\tilde{F} \right] \nonumber\\
&-&\sum_{j=x,y,z}\epsilon_{ajc}q_j q_b \frac{1}{6}\Big[\frac{e\omega v_F^3 \tau^3}{c} \left(1+2i\omega\tau\right)\tilde{V}_{\lambda}^{(1)} C_2 +\frac{v_F \hbar^2 \omega \tau}{2e} \left(\tilde{V}_{\lambda}^{(1)}\right)^{*} \left(1-i\omega\tau\right)F
\nonumber\\
&+& i\frac{e\hbar^2}{c} \left(\tilde{V}_{\lambda}^{(1)} +2\left(\tilde{V}_{\lambda}^{(1)}\right)^{*}\right)\tilde{F}
\Big]
\Bigg\} +c.c.,
\end{eqnarray}
and
\begin{eqnarray}
\label{CKT-Second-gamma-tens-0}
\gamma_{abc} &=& -\frac{e}{4}\sum_{\lambda=\pm} q_a \delta_{bc} \frac{e^2\hbar^2 v_F^3 \omega \tau^2}{60c^2}\left[ i\frac{\omega \tau}{2} \left(1+2i\omega\tau\right) -\left(1+3i\omega\tau\right)
\right]F +c.c.
\end{eqnarray}
In addition, we used
\begin{equation}
\label{CKT-Second-tV-def}
\tilde{V}_{\lambda}^{(1)} = \frac{v_F^2 \tau e}{3\omega} \frac{1-2(\omega \tau)^2+2i\omega\tau}{1-(\omega \tau)^2+i\omega \tau -(v_Fq\tau)^2/3}.
\end{equation}

Notice that the acoustic frequency is quite small and it is legitimate to expand is small $\omega \tau$. In this regime, we can approximate the above response functions as
\begin{eqnarray}
\label{CKT-Second-sigma-tens}
\sigma_{abc}
&\approx& -q_a \delta_{bc} \frac{4e^3v_F^3 \tau^4 \omega}{15} C_2
+q_b \delta_{ac} \frac{e^3v_F^3 \tau^2}{9\omega} C_2
-q_aq_bq_c \frac{e^3 v_F^5 \tau^4}{135\omega} C_2
-\delta_{ab} q_c \frac{2e^3 v_F^5 \tau^4 q^2}{45\omega} C_2
+{\cal O}[(\omega\tau)^3],
\\
\label{CKT-Second-kappa-tens}
\kappa_{abc} &\approx& \epsilon_{abc} \frac{e^3}{12c} \left(2v_F^3 \tau^2 C_2-6\hbar^2 \tilde{F} \right)
+\sum_{j=x,y,z}\epsilon_{ajc}q_j q_b \frac{e^3}{18c} \left(v_F^5 \tau^4 C_2-3\hbar^2 v_F^2 \tau^2 \tilde{F} +\frac{\hbar^2 v_F^2 \tau^2 F}{2}\right)
+{\cal O}[(\omega\tau)^2],\\
\label{CKT-Second-gamma-tens}
\gamma_{abc} &\approx&
q_a \delta_{bc} \frac{e^3\hbar^2 v_F^3 \tau^2 \omega}{60 c^2} F
+{\cal O}[(\omega\tau)^2],
\end{eqnarray}
where one should also set $\mu_5\to0$. In addition to the shorthand notations in Eqs.~(\ref{CKT-First-C1})--(\ref{CKT-First-C3}), we introduced
\begin{eqnarray}
\label{CKT-Second-C4}
F &=& \sum_{\eta=\pm} \int\frac{d^3p}{(2\pi \hbar)^3} \frac{1}{p^3}\partial_{\epsilon_{\mathbf{p}}} f^{\rm eq}_{\lambda} =
-\frac{1}{2\pi^2 \hbar^3 T} F_0\left(\frac{\mu_{\lambda}}{T}\right),\\
\label{CKT-Second-C5}
\tilde{F} &=& \sum_{\eta=\pm} \eta\int\frac{d^3p}{(2\pi \hbar)^3} \frac{1}{p^4} f^{\rm eq}_{\lambda} =
\frac{1}{2\pi^2\hbar^3} \int_{\Lambda_{\rm IR}}^{\infty} \frac{dp}{p^2} f_{\lambda}^{\rm eq} = \frac{1}{2\pi^2\hbar^3} \sum_{\eta=\pm}\eta\left[\frac{1}{\Lambda_{\rm IR}} \frac{1}{1+e^{(v_F\Lambda_{\rm IR}-\mu_{\lambda})/T}}
+ \eta v_F \int_{\Lambda_{\rm IR}}^{\infty} \frac{dp}{p} \partial_{\epsilon_{\mathbf{p}}} f^{\rm eq}_{\lambda}
\right] \nonumber\\
&=&\frac{v_F}{2\pi^2\hbar^3 T} \tilde{F}_0\left(\frac{\mu_{\lambda}}{T}, \frac{v_F \Lambda_{\rm IR}}{T}\right).
\end{eqnarray}

\begin{figure*}[!ht]
\begin{center}
\includegraphics[width=0.4\textwidth]{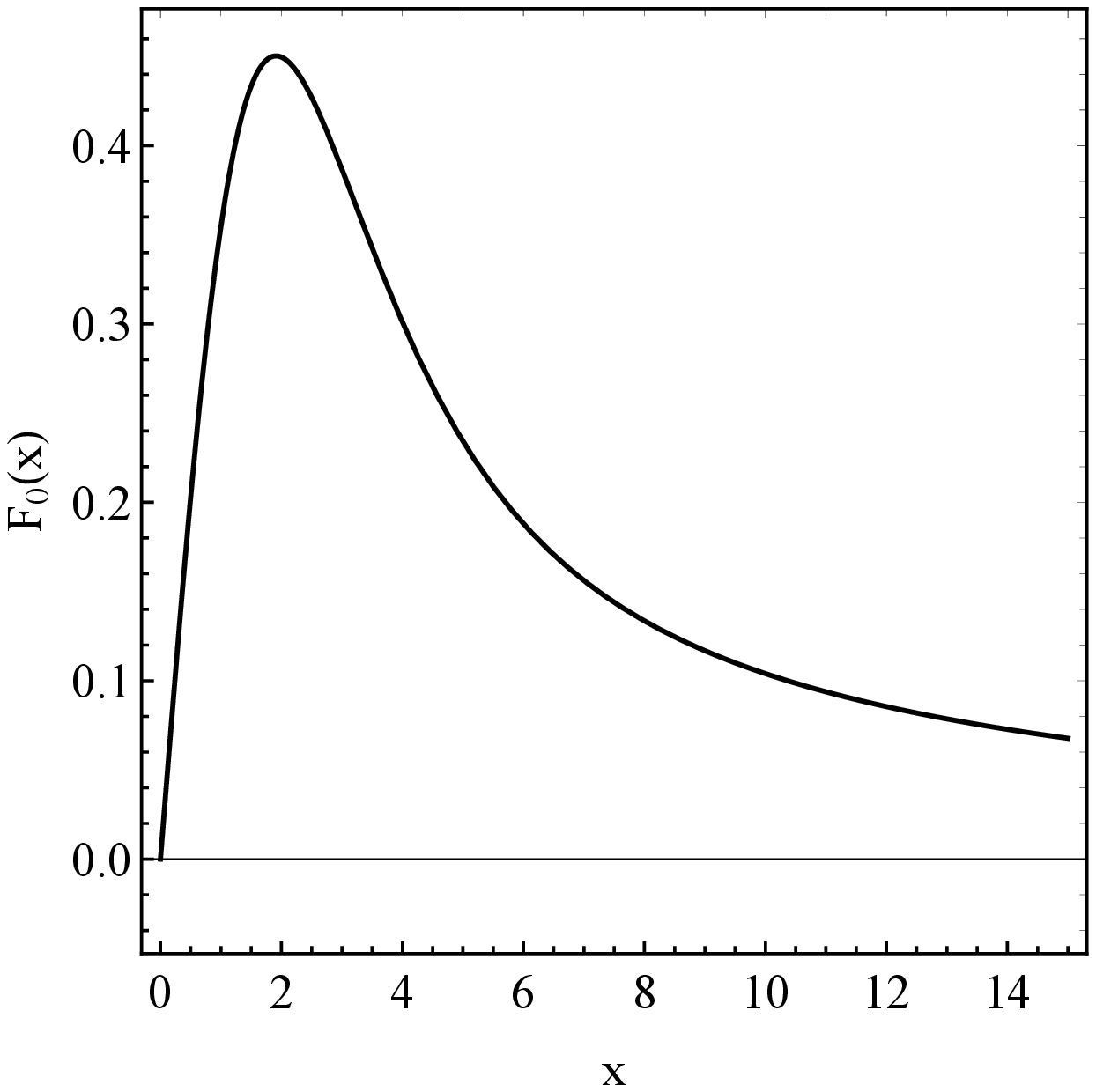}\hfill
\includegraphics[width=0.4\textwidth]{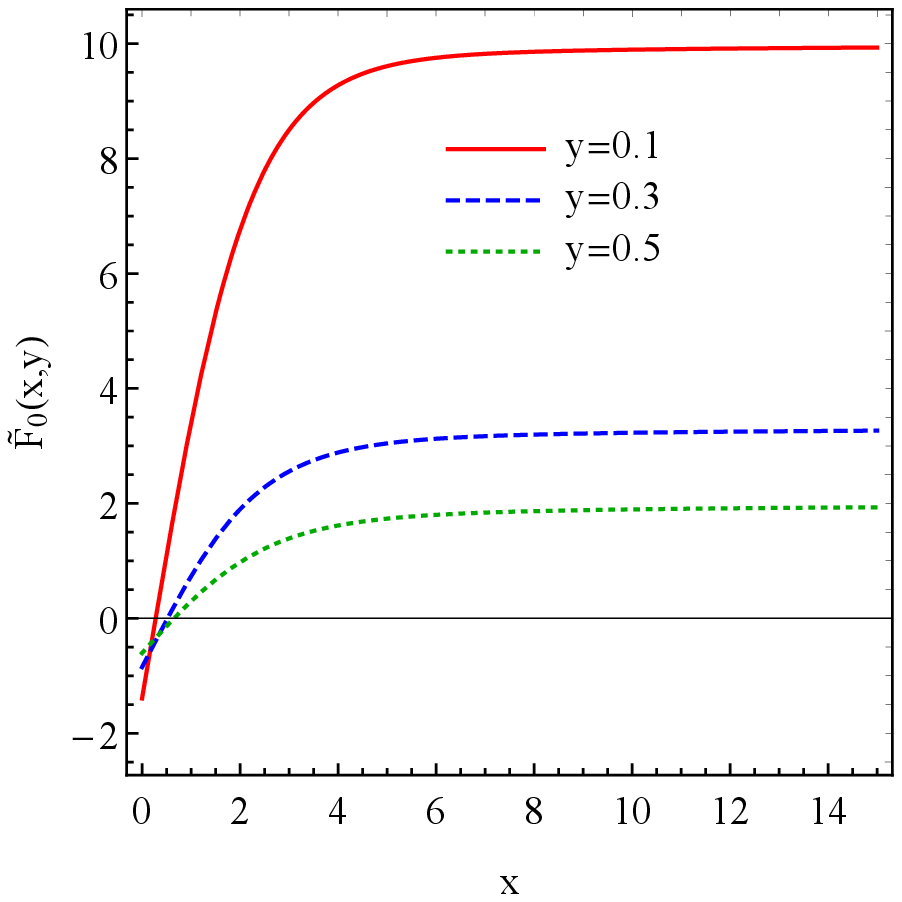}
\end{center}
\caption{The dependency of $F_0\left(x\right)$ (left panel) and $\tilde{F}_0\left(x,y\right)$ (right panel) on $x$. In the case of $\tilde{F}_0\left(x,y\right)$, we set $y=0.1$ (red solid line), $y=0.3$ (blue dashed line), and $y=0.5$ (green dotted line).
}
\label{fig:CKT-second-tC4}
\end{figure*}

We present the function $F_0\left(x\right)$ in the left panel of Fig.~\ref{fig:CKT-second-tC4}. High- and low-temperature asymptotes of $F_0\left(x\right)$ equal
$F_0\left(x\right)\simeq 7 \zeta(3) x/(2\pi^2)\approx 0.426  x$ for $x\to0$ and
$F_0\left(x\right)\simeq x^{-1}$ for $x\to\infty$, respectively. The function $F_0\left(x\right)$ could be approximated by the Pad\'e approximant of order [5/6] as
\begin{eqnarray}
\label{CKT-Second-tC4-Pade}
F_0\left(x\right) \simeq \frac{7\zeta(3)}{2\pi^2}  \frac{x +0.03533x^3 +0.0007432x^5}
{1+0.2290x^2+0.01567 x^4+0.0003098x^6}.
\end{eqnarray}
The function $\tilde{F}_0\left(x,y\right)$ is presented in the right panel of Fig.~\ref{fig:CKT-second-tC4} for a few values of $y$. It is clear that it has a $1/y$ dependence and quickly reaches a constant value at large values of $x$.
In our numerical calculations, we introduced the infrared cutoff $\Lambda_{\rm IR} \simeq \omega/v_s$, which is of order of the sound wavelength. One can alternatively use the magnitude of the oscillating pseudomagnetic field $\Lambda_{\rm IR} \simeq \sqrt{\hbar |eB_{5,0}|/c}$. However, the latter approach does not allow to define the susceptibility.
Such a cutoff separates the phase space of large momenta, where the semiclassical description provided by the second order chiral kinetic theory is valid, from the infrared region $p<\Lambda_{\rm IR}$, where such a description fails (for details, see also the discussion in Ref.~[S3]
). We believe that the appearance of such divergences is an artifact of the expansion in the second order chiral kinetic theory, which is related to the terms $\bm{\Omega}_{\lambda}^{(1)}$, $\epsilon_{\mathbf{p}}^{(2)}$, and $\mathbf{v}_{\mathbf{p}}^{(2)}$ and
is manifested for certain Hall-like responses $\sim  \mathbf{E}_{\lambda}\times \mathbf{B}_{\lambda}$.
On the other hand, since the presence of terms $\sim \Lambda_{\rm IR}^{-1}$ does not affect our qualitative conclusion regarding the possibility of the acoustogalvanic rectification as well as does not affect the longitudinal response, we leave the investigation of a proper treatment of such terms for future studies. For example, one can employ the Kubo approach or the Wigner function formalism~[S24],
where there expansion in terms of the terms of small  $\sim\bm{\Omega}_{\lambda}\cdot\mathbf{B}_{\lambda}$ is not performed.

\subsection{C. Acoustogalvanic susceptibility}
\label{sec:charge-current-AG}

Since the pseudo-electromagnetic fields $\mathbf{E}_5$ and $\mathbf{B}_5$ are secondary fields induced by the displacement vector $\mathbf{u}$, it is convenient to rewrite the electric current (\ref{CKT-Second-j-DC-1}) in terms of these fields as
\begin{equation}
\label{App-j-DC-u}
j^{\rm dc}_{a} = \chi^{\rm AG}_{abc} u_b u^{*}_c,
\end{equation}
where $ \chi^{\rm AG}_{abc} $ is defined as the acoustogalvanic susceptibility.
By using Eq.~(\ref{Model-A5-vec}), we derive the following component of the acoustogalvanic response function $\chi^{\rm AG}_{zzz}$
\begin{equation}
\label{App-cc-second-chi-AG-zzz}
\chi^{\rm AG}_{zzz} = \frac{\omega^4}{v^2_s} \frac{\hbar^2 b^2}{e^2} \left[\beta+ \tilde\beta(b)\right]^2  \sigma_{zzz}.
\end{equation}
The corresponding components of the acoustogalvanic tensor $\chi^{\rm AG}_{xxx}=\chi^{\rm \sigma}_{xxx}+\chi^{\rm \kappa}_{xxx}+\chi^{\rm \gamma}_{xxx}$ for small $\omega\tau$ are
\begin{eqnarray}
\label{cc-second-chi-AG-xxx-sigma}
\chi^{\sigma}_{xxx} &=& \frac{e\mu v_F b^2\tilde{\beta}^2 \tau^4 \omega^6}{30\pi^2 \hbar v_s^3},\\
\label{cc-second-chi-AG-xxx-kappa}
\chi^{\kappa}_{xxx} &=& -\frac{c\hbar^2 b^2\tilde{\beta}^2 \omega^4}{2e^2v_s^3} G_1\left(\mu, T,\Lambda_{\rm IR}\right) -\frac{e\mu v_F \tilde{\beta}^2 \tau^2 \omega^4}{12\pi^2 \hbar v_s^3},\\
\label{cc-second-chi-AG-xxx-gamma}
\chi^{\gamma}_{xxx} &=& \frac{\hbar^2 c^2 b^2\tilde{\beta}^2 \tau^2 \omega^6}{e^2 v_s^4}G_2\left(\mu, T\right).
\end{eqnarray}
Functions $G_1\left(\mu, T,\Lambda_{\rm IR}\right)$ and $G_2\left(\mu, T\right)$ (see also Eqs.~(13) and (14) in the main text) are related to the functions $F_0$ and $\tilde{F}_0$ defined in Eqs.~(\ref{CKT-Second-C4}) and (\ref{CKT-Second-C5}) as
\begin{eqnarray}
\label{CKT-Second-G1}
G_1\left(\mu, T,\Lambda_{\rm IR}\right) &=& \frac{e^3 v_F}{2\pi^2\hbar cT} \tilde{F}_0,\\
\label{CKT-Second-G2}
G_2\left(\mu, T\right) &=& -\frac{e^3 v_F^3}{120\pi^2 c^2 v_s \hbar T} F_0.
\end{eqnarray}

\subsection{D. Acoustogalvanic chiral current}
\label{sec:charge-current-AG-5}

For the sake of completeness, let us also consider the acoustogalvanic chiral current density $\mathbf{j}_5^{\rm dc} = \sum_{\lambda=\pm}\lambda \mathbf{j}_{\lambda}^{(2)}$:
\begin{equation}
\label{CKT-Second-j5-DC-1}
j^{\rm dc}_{5,a} = \sigma_{5, abc} E_{5,b} E^{*}_{5,c} +\kappa_{5, abc} \frac{1}{2}\left(E_{5,b} B^{*}_{5,c}+E_{5,b}^{*} B_{5,c}\right) +\gamma_{5, abc} B_{5,b} B^{*}_{5,c}.
\end{equation}
By using the results from the previous Subsection (see Eqs.~(\ref{CKT-Second-sigma-tens}), (\ref{CKT-Second-kappa-tens}), and (\ref{CKT-Second-gamma-tens})), we obtained the following chiral response tensors in the limit of small $\omega \tau$ and $v_s/v_F$:
\begin{eqnarray}
\label{CKT-Second-sigma5-tens}
\sigma_{5, abc} &\approx& -\sum_{j=x,y,z}\epsilon_{ajb}q_j q_c \frac{e^3\hbar v_F^3 \tau^3}{18} C_3 \left(1-\omega^2\tau^2\right) +{\cal O}[(\omega\tau)^3], \\
\label{CKT-Second-kappa5-tens}
\kappa_{5, abc} &\approx& \delta_{ac} q_b \frac{e^3\hbar v_F^3 \tau}{6c\omega} \left(1+ \frac{v_F^2\omega^2\tau^2}{3v_s^2} \right) C_3 +{\cal O}[(\omega\tau)^3], \\
\label{CKT-Second-gamma5-tens}
\gamma_{5, abc} &\approx& {\cal O}[(\omega\tau)^3].
\end{eqnarray}

\begin{figure}[t]
\begin{center}
\includegraphics[width=0.365\textwidth]{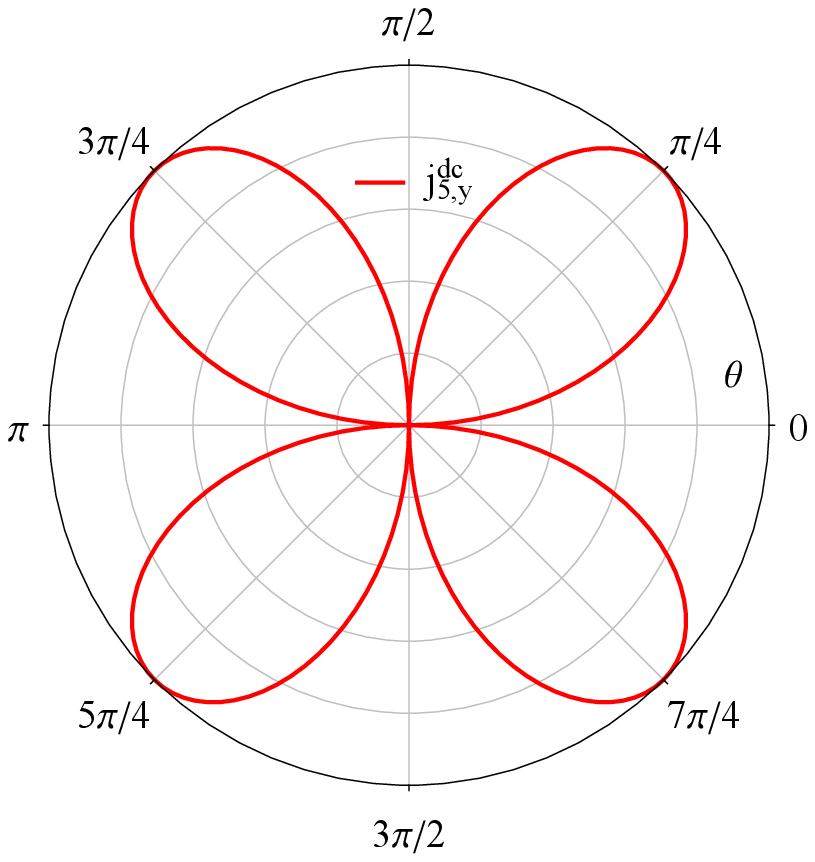}
\end{center}
\caption{The dependency of the rectified chiral current component $j_{5,y}^{\rm dc}$ on the angle $\theta$ between the chiral shift $\mathbf{b}$ and the wave vector $\mathbf{q}$. We fixed $\omega=1~\mbox{MHz}$. A typical value of the rectified current is very small, reaching $j^{\rm dc}\approx 29\times10^{-5}~u_0^2/{\rm \mu m^2}~[{\rm n A/{\rm cm}^2]}$ at $\theta=\pi/4$.}
\label{fig:CKT-second-j5}
\end{figure}

We present the angular dependence of the acoustogalvanic chiral current $j_{5,y}^{\rm dc}$ in Fig.~\ref{fig:CKT-second-j5}. Note that $j_{5,x}^{\rm dc}=j_{5,z}^{\rm dc}=0$. Therefore, unlike the case of the rectified electric current, the chiral one is not coplanar with $\mathbf{q}$ and $\mathbf{b}$. Physically, chiral current corresponds to the spin polarization of a sample. As in the main text, we used the numerical parameters valid for the Dirac semimetal Cd$_3$As$_2$~[S19--S23]:
$v_{\rm F}\approx 1.5\times 10^8~{\rm cm/s}$, $\mu\approx 200~{\rm meV}$, $b\approx 1.6~{\rm nm}^{-1}$, $ v_s \approx 2.3\times 10^5~{\rm cm/s}$, and $\tau\approx~1~{\rm ps}$. In addition, we assume that $\beta\approx 1$, $T=5~{\rm K}$, and $\tilde{\beta}(b)\approx 1$. For the parameters used, the magnitude of the current is small and reaches $I_5^{\rm dc}\sim 10^{-6}~{\rm nA}$ for mm-sized devices and $u_0=10\,\mbox{nm}$.

\section{S III. Time-reversal symmetric case}
\label{sec:TR-symm}

In this Section, we qualitatively address the case of time-reversal ($\mathcal{T}$) symmetric Weyl semimetals, where, however, the parity ($\mathcal{P}$) symmetry is broken. We focus on the simplest case, where the energy separation between the Weyl nodes is $2b_0$ and there is only $A_{0,5}$ component of the axial gauge field given in Eq.~(\ref{Model-A5-0}). By using the approach discussed in Sec.~\ref{sec:charge-current}, where, however, the pseudomagnetic field $\mathbf{B}_5$ is absent, we derive the following acoustogalvanic current:
\begin{equation}
\label{TR-symm-j-DC}
j^{\rm dc}_a = \sigma_{abc} E_{5,b} E^{*}_{5,c}
\end{equation}
and its chiral counterpart
\begin{equation}
\label{TR-symm-j5-DC}
j^{\rm dc}_{5,a} = \sigma_{5, abc} E_{5,b} E^{*}_{5,c}.
\end{equation}
The response tensors are $\sigma_{abc}=\sum_{\lambda=\pm}\sigma_{\lambda, abc}$ and $\sigma_{5, abc}=\sum_{\lambda=\pm}\lambda\sigma_{\lambda, abc}$ where
\begin{eqnarray}
\label{TR-symm-sigma-tens}
\sigma_{\lambda, abc} &=& -\frac{e}{4} \Bigg\{
-q_a \delta_{bc} \frac{ie^2v_F^3 \tau^3}{30} \left(1+2i\omega\tau\right) C_2
-\delta_{ac}q_b \frac{e\tau v_F^2}{3}\left[ \frac{4ev_F \tau^3 \omega}{5} - i\omega \tau \tilde{V}_{\lambda}^{(1)} \left(1+i\omega\tau\right)
+\tilde{V}_{\lambda}^{(1)} \right] C_2 \nonumber\\
&+&q_a q_b q_c \frac{\tau v_F^2}{3} \tilde{V}_{\lambda}^{(1)}\left[ \frac{2ev_F \tau^2}{5} \left(1+2i\omega\tau\right) + \omega \tau \tilde{V}_{\lambda}^{(1)} \left(1+i\omega\tau\right) +i\tilde{V}_{\lambda}^{(1)} \right] C_2 \nonumber\\
&+&\delta_{ab}q_c q^2 \left(\tilde{V}_{\lambda}^{(1)}\right)^{*} \frac{2ev_F^3 \tau^3}{15} \left(1+2i\omega\tau\right) C_2
+\epsilon_{ajb}q_j q_c \left(\tilde{V}_{\lambda}^{(1)}\right)^{*} \frac{i\lambda ev_F \hbar \tau}{6} \left[1-2(\omega \tau)^2 \right] C_3
\Bigg\} +c.c..
\end{eqnarray}
Coefficients $C_1$ and $C_3$ are given in Eqs.~(\ref{CKT-First-C1}) and (\ref{CKT-First-C3}), respectively. Note also that $\mu_5=b_0$ in these equations and, for qualitative estimates, we neglected the effects of dynamical electromagnetism.

In the case of small frequencies and $v_s/v_F\ll1$, we have
\begin{eqnarray}
\label{TR-symm-sigma-tens-1}
\sigma_{abc} &\approx&
\frac{e^3\tau^2 v_F \mu}{2\pi^2 \hbar^3 \omega} \left\{q_a \delta_{bc} \frac{\tau^2 \omega^2}{10}
-q_b \delta_{ac} \frac{1}{6}
+q_aq_bq_c \frac{v_F^2\tau^2}{90}
-\delta_{ab} q_c \frac{v_F^2q^2 \tau^2}{15} \right\}
+{\cal O}[(\omega\tau)^3], \\
\label{TR-symm-sigma-tens-2}
\sigma_{5,abc} &\approx& \frac{e^3\tau^2 v_F \mu_5}{2\pi^2 \hbar^3 \omega} \left\{q_a \delta_{bc} \frac{\tau^2 \omega^2}{10}
-q_b \delta_{ac} \frac{1}{6}
+q_aq_bq_c \frac{v_F^2\tau^2}{90}
-\delta_{ab} q_c \frac{v_F^2q^2 \tau^2}{15} \right\}
+\sum_{j=x,y,z}\epsilon_{ajb}q_j q_c \frac{e^3v_F^2 \tau^3}{36\pi^2 \hbar^2} \left(1-\omega^2\tau^2\right) \nonumber\\
&+&{\cal O}[(\omega\tau)^3].
\end{eqnarray}

The corresponding acoustogalvanic electric and chiral susceptibilities read as
\begin{eqnarray}
\label{TR-symm-chi}
\chi_{abc} &=& -\delta_{ab}\delta_{ac} \frac{e b_0^2\beta^2 v_F \mu\tau^2 \omega^4}{18\pi^2 \hbar^3 v_s^5}
+{\cal O}[(\omega\tau)^3],\\
\label{TR-symm-chi5}
\chi_{5,abc} &=& -\delta_{ab}\delta_{ac} \frac{e b_0^2\beta^2 v_F\mu_5 \tau^2 \omega^4}{18\pi^2 \hbar^3 v_s^5}
+{\cal O}[(\omega\tau)^3],
\end{eqnarray}
respectively.

Since there is no preferred direction in the system, both electric and chiral rectified currents are isotropic. Their direction is determined by the sound wave vector. Assuming $b_0=10\,\mbox{meV}$ and using parameters given at the end of Sec.~\ref{sec:charge-current-second}, we estimate the electric and chiral current densities as $j^{\rm dc}\approx 230~u_0^2/{\rm \mu m^2}~[{\rm \mu A/{\rm cm}^2]}$ and $j^{\rm dc}_5\approx 11~u_0^2/{\rm \mu m^2}~[{\rm \mu A/{\rm cm}^2]}$, respectively. As one can see, even for small energy separation, the acoustogalvanic currents might be noticeably larger than in the case of $\mathcal{T}$ symmetry broken Weyl semimetals. This can be attributed to the fact that the corresponding pseudoelectric field is determined by the spatial rather than the time derivative from the gauge field. Therefore, due to the smallness of the sound velocity $v_s$, the contribution to the pseudoelectric field from the scalar part of the axial gauge potential could be larger than that from the vector one, i.e.,  $c|\partial_{\mathbf{r}}A_{0,5}|/|\partial_{t}\mathbf{A}_{5}| \propto b_0/(v_s \hbar b)$ even for sufficiently large separation of the Weyl nodes in momentum space. In addition, the chiral current is comparable to the electric one. Thus, the simple case of Weyl semimetals with preserved $\mathcal{T}$ symmetry but broken $\mathcal{P}$ one is also characterized by the acoustogalvanic effect, where, however, the angular distribution of the current lacks preferred direction inherent for $\mathcal{T}$ symmetry broken models.

\section{S IV. Strain effects in type-II Weyl semimetals: a lattice model}
\label{sec:type-II}

In this Section, we address the possibility to describe strains in terms of effective gauge fields in type-II Weyl semimetals~[S25].
We employ a four-band tight-binding model used in Refs.~[S9,S10,S26--S29],
where spin-up and spin-down electrons of $s$ and $p$ atomic orbitals hop on a cubic lattice. By adding the tilt term ${\cal H}_{\rm tilt}$ to the Hamiltonian, we obtain
\begin{align}
\label{type-II-H}
{\cal H} =
\begin{bmatrix}
h_{11} &  h_{12} \\ h^\dagger_{12} & h_{22}
\end{bmatrix} + {\cal H}_{\rm tilt},
\end{align}
where $\bm{\sigma}$ is the vector of Pauli matrices as well as $\mathds{1}_2$ and $\mathds{1}_4$ are $2\times2$ and $4\times4$ unit matrices, respectively. Further, we have the following components:
\begin{align}
&h_{11} = \hbar v_{\rm F} \sum_{i=x,y} \sigma_i
\Big[ (1-\beta_t u_{ii})  \xi_i  +\beta^{\prime}_t \sum_{j\neq i} u_{ij} \xi_j \Big]
- \sigma_z (b_3-{\cal M})
 \\
& h_{22} =- \hbar v_{\rm F} \sum_{i=x,y} \sigma_i
\Big\{ (1-\beta_t u_{ii})  \xi_i +\beta^{\prime}_t \sum_{j\neq i} u_{ij} \xi_j \Big\}
-  \sigma_z (b_3+{\cal M})
\\
& h_{12} =-\hbar v_{\rm F} \mathds{1}_{2} \Big\{ (1-\beta_t u_{zz})  \xi_z +   \sum_{j\neq z} u_{zj} \xi_j\Big\}
\\
&{\cal H}_{\rm tilt} = \hbar v_{\rm F} \mathds{1}_{4} \alpha  (1-\beta_f u_{zz}) \zeta_z,
\end{align}
where ${\cal M}=m+3r - r\sum_{i=x,y,z}(1-\beta_r u_{ii}) \zeta_i$, $\xi_i=\sin{(k_i a)}$, $\zeta_i=\cos{(k_i a)}$, and $a$ is the lattice constant. Note that $v_{\rm F}$ is the Fermi velocity and $\alpha$ is the dimensionless tilt parameter. The parameters $\beta_t,\beta'_t,\beta_r,\beta_f$ are the Gr\"uneisen parameters for different hopping channels.

In the absence of strain and tilt, Hamiltonian (\ref{type-II-H}) describes a Weyl semimetal with two Weyl nodes located at $k_z^{(\pm)} \approx \pm \sqrt{(b_3^2-m^2)/(\hbar^2v_F^2 +a^2rm)}$. Type-II Weyl semimetal is realized for $\alpha>1/[1+(\beta_t-\beta_f)u_{zz}]$.

\begin{figure*}[!ht]
\begin{center}
\includegraphics[width=0.32\textwidth]{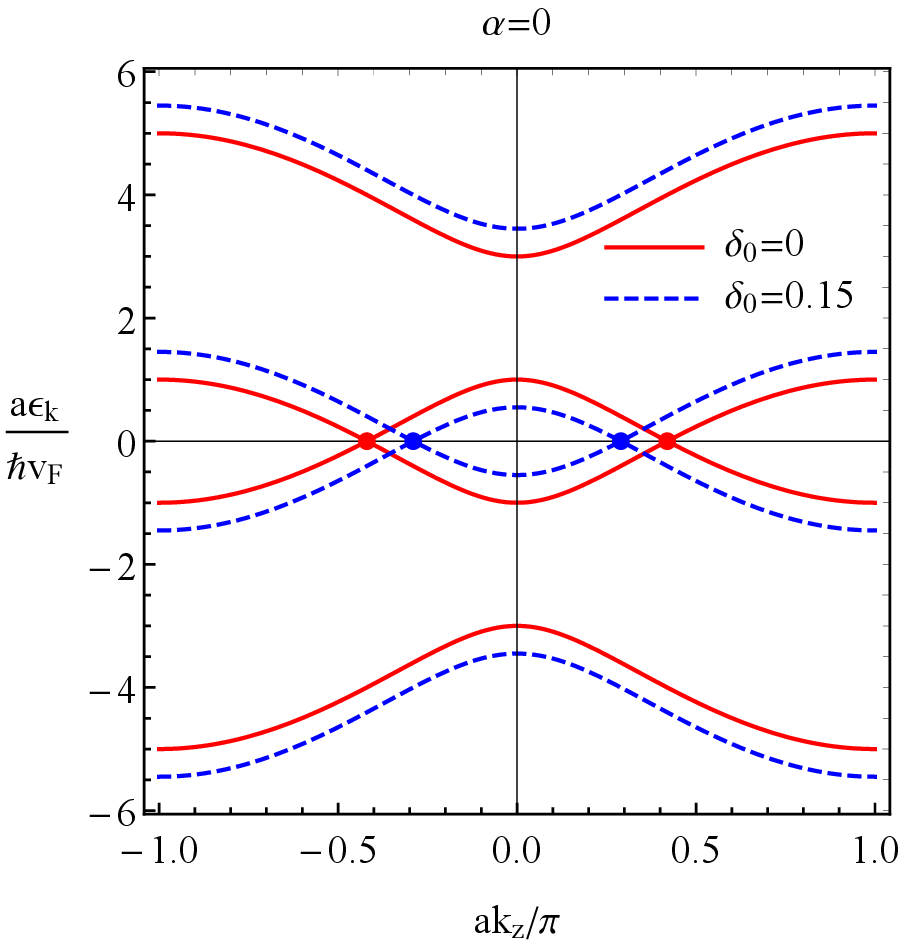}
\hfill
\includegraphics[width=0.32\textwidth]{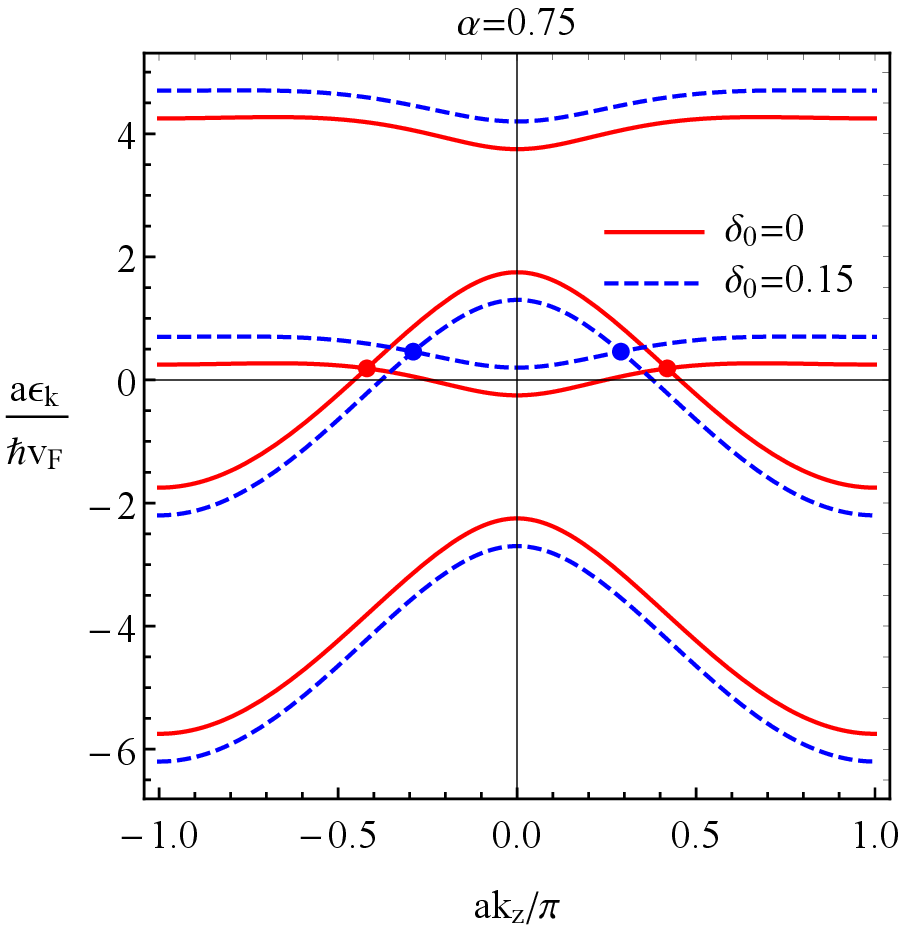}
\hfill
\includegraphics[width=0.32\textwidth]{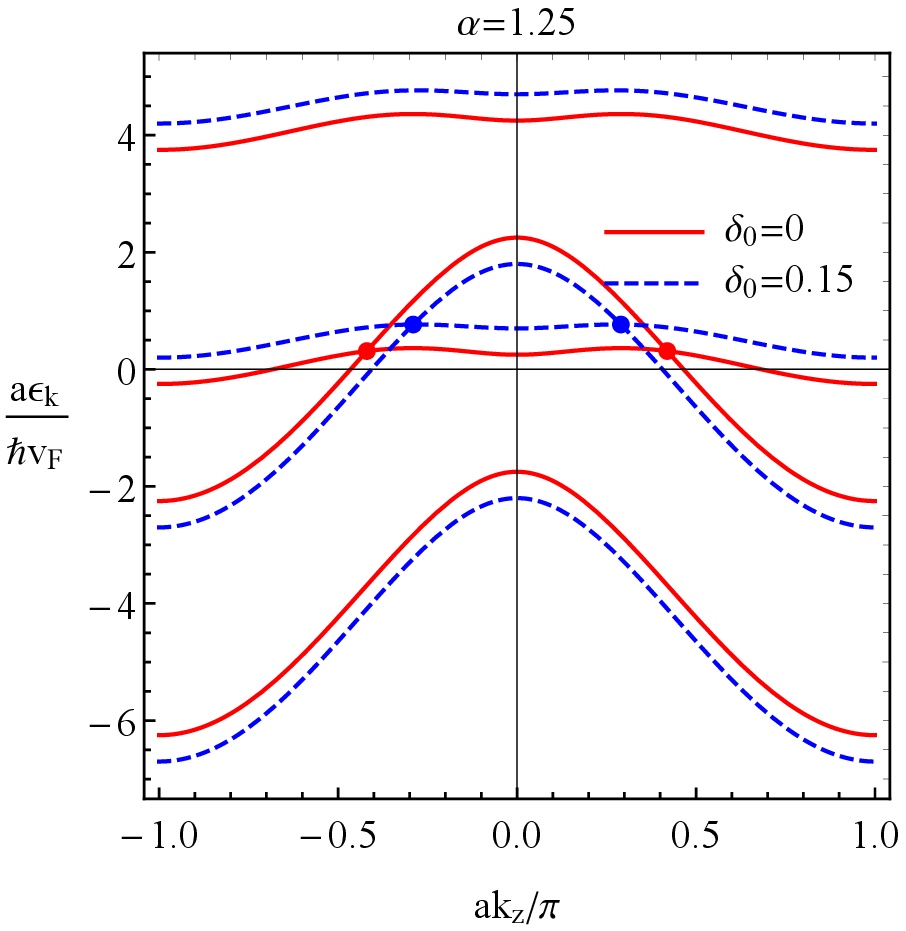}
\end{center}
\hspace{0.01\textwidth}{\small (a)}\hspace{0.32\textwidth}{\small (b)}\hspace{0.32\textwidth}{\small (c)}\\[0pt]
\caption{The energy spectrum of Hamiltonian (\ref{type-II-H}) without (red solid lines) and with stretching $u_{ij}=\delta_0 \delta_{ix}\delta_{jx}$ (blue dashed lines) for $\alpha=0$ (panel (a)), $\alpha=0.75$ (panel (b)), and $\alpha=1.25$ (panel (c)). We used  $\beta_t=\beta^{\prime}_t=\beta_r=\beta_f=3 \hbar v_F/a$, $r=m=\hbar v_F/a$, $b_3=2 \hbar v_F/a$, and $\delta_0=0.15$.
}
\label{fig:type-II-spectrum}
\end{figure*}

We plot the energy spectrum of Hamiltonian (\ref{type-II-H}) at a few values of tilt parameter $\alpha$ in Fig.~\ref{fig:type-II-spectrum} for a uniform strain along the $x$ axis, $u_{ij}=\delta_0 \delta_{ix}\delta_{jx}$. As one can see, strain shifts the position of Weyl nodes in momentum ($\alpha=0$) as well as both in momentum and energy for $\alpha\neq0$. Therefore, as in the case of simple type-I Weyl semimetals considered in the main text and Sec.~\ref{sec:Model-general}, strains can be interpreted in terms of effective gauge fields in tilted and overtilted (type-II) Weyl semimetals. Then, the acoustogalvanic effect originating from the dynamical axial gauge fields should also occur in such systems.

Among the key qualitative features that distinguish type-I and type-II Weyl semimetals, we mention a strong anisotropy defined by the tilt as well as the formation of the electron and hole pockets. Since both sound wave vector and tilt break the $\mathcal{P}$ symmetry, we expect several new terms in the acoustogalvanic response where these vectors appear interchangeably. Detailed analysis of the acoustogalvanic effect in tilted Weyl semimetals, especially in the type-II case, requires a more rigorous treatment and will be reported elsewhere.

\section{S V. Useful formulas}
\label{sec:App-formulas}

In this section, several useful formulas related to the integration over momenta and angles are presented.
By making use of the short-hand notation for the global equilibrium Fermi--Dirac distribution function $f^{\rm eq}_{\lambda} =1/[e^{(v_F p-\mu_{\lambda})/T}+1]$ at $\eta=+$, it is straightforward to derive the following formulas:
\begin{eqnarray}
\label{App-1-integral-4a}
\int\frac{d^3p}{(2\pi \hbar)^3} p^{n-2} F(\theta) f^{\rm eq}_{\lambda}
&=& -\frac{T^{n+1} \Gamma(n+1) }{4\pi^2 \hbar^3 v_F^{n+1}}  \mbox{Li}_{n+1}\left(-e^{\mu_{\lambda}/T}\right) \int_{-1}^{1}d\cos{\theta}\, F(\theta),\\
\label{App-1-integral-4b}
\int\frac{d^3p}{(2\pi \hbar)^3} p^{n-2} F(\theta) \frac{\partial f^{\rm eq}_{\lambda}}{\partial p}
&=& \frac{T^{n} \Gamma(n+1)}{4\pi^2 \hbar^3 v_F^{n}}  \mbox{Li}_{n}\left(-e^{\mu_{\lambda}/T}\right) \int_{-1}^{1}d\cos{\theta}\, F(\theta),\\
\label{App-1-integral-4c}
\int\frac{d^3p}{(2\pi \hbar)^3} p^{n-2} F(\theta) \frac{\partial^2 f^{(\rm eq)}_{\lambda}}{\partial p^2}
&=& -\frac{T^{n-1} \Gamma(n+1)}{4\pi^2 \hbar^3 v_F^{n-1}}   \mbox{Li}_{n-1}\left(-e^{\mu_{\lambda}/T}\right) \int_{-1}^{1}d\cos{\theta}\, F(\theta),
\end{eqnarray}
where $T\partial f^{\rm eq}_{\lambda}/\partial p =-v_F T\partial f^{\rm eq}_{\lambda}/\partial \mu_{\lambda} =-v_F e^{(v_F p -\mu_{\lambda})/T}/[e^{(v_F p-\mu_{\lambda})/T}+1]^2$, $n\geq0$, $\mbox{Li}_{n}(x)$ is the polylogarithm function, and $F(\theta)$ is a function that depends only on the polar angle $\theta$. The polylogarithm functions at $n=0,1$ can be rewritten in terms of the elementary functions
\begin{eqnarray}
\mbox{Li}_{0}\left(-e^{x}\right) &=& -\frac{1}{1+e^{-x}}, \\
\mbox{Li}_{1}\left(-e^{x}\right) &=& -\ln{\left(1+e^{x}\right)}.
\label{App-1-App-polylog}
\end{eqnarray}
The following identities are useful when summing over $\eta=\pm$:
\begin{eqnarray}
\mbox{Li}_{0}\left(-e^{x}\right)+\mbox{Li}_{0}\left(-e^{-x}\right) &=& -1,\\
\mbox{Li}_{1}\left(-e^{x}\right)-\mbox{Li}_{1}\left(-e^{-x}\right) &=& -x,\\
\mbox{Li}_{2} (-e^{x}) +\mbox{Li}_{2} (-e^{-x})   &=& -\frac{1}{2}\left(x^2+\frac{\pi^2}{3}\right).
\end{eqnarray}

Finally, by integrating over the angular coordinates, one can derive the following general relations:
\begin{eqnarray}
&&\int \frac{d^3 p}{(2\pi\hbar)^3}  \mathbf{p}  f(p^2) = 0,\\
&&\int \frac{d^3 p}{(2\pi\hbar)^3}  \mathbf{p} (\mathbf{p}\cdot \mathbf{a}) f(p^2) = \frac{\mathbf{a}}{3} \int \frac{d^3 p}{(2\pi\hbar)^3}  p^2 f(p^2),\\
&&\int \frac{d^3 p}{(2\pi\hbar)^3}  \mathbf{p} (\mathbf{p}\cdot \mathbf{a}) (\mathbf{p}\cdot \mathbf{b})  f(p^2) = 0,\\
&&\int \frac{d^3 p}{(2\pi\hbar)^3}  \mathbf{p} (\mathbf{p}\cdot \mathbf{a}) (\mathbf{p}\cdot \mathbf{b}) (\mathbf{p}\cdot \mathbf{c}) f(p^2) = \frac{\mathbf{a}(\mathbf{b}\cdot\mathbf{c}) +\mathbf{b}(\mathbf{a}\cdot\mathbf{c}) +\mathbf{c}(\mathbf{a}\cdot\mathbf{b})}{15} \int \frac{d^3 p}{(2\pi\hbar)^3}  p^2 f(p^2).
\end{eqnarray}

\vspace{0.5cm}
\begin{center}
\noindent\rule{6cm}{1pt}
\end{center}
\vspace{0.5cm}

\begin{itemize}

\item[[S1\!\!]]
D.~Xiao, M.-C.~Chang, and Q.~Niu,
    \href{https://doi.org/10.1103/RevModPhys.82.1959}{Rev. Mod. Phys. {\bf 82}, 1959 (2010).}

\item[[S2\!\!]]
D.~T.~Son and N.~Yamamoto,
    \href{https://doi.org/10.1103/PhysRevD.87.085016}{Phys. Rev. D {\bf 87}, 085016 (2013).}

\item[[S3\!\!]]
M.~A.~Stephanov and Y.~Yin,
    \href{https://doi.org/10.1103/PhysRevLett.109.162001}{Phys. Rev. Lett. {\bf 109}, 162001 (2012).}

\item[[S4\!\!]]
D.~T.~Son and N.~Yamamoto,
    \href{https://doi.org/10.1103/PhysRevLett.109.181602}{Phys. Rev. Lett. {\bf 109}, 181602 (2012).}

\item[[S5\!\!]]
D.~T.~Son and B.~Z.~Spivak,
    \href{https://doi.org/10.1103/PhysRevB.88.104412}{Phys. Rev. B {\bf 88}, 104412 (2013).}

\item[[S6\!\!]]
    Y.~Gao, S.~A.~Yang, and Q.~Niu,
    \href{https://doi.org/10.1103/PhysRevLett.112.166601}{Phys. Rev. Lett. {\bf 112}, 166601 (2014).}

\item[[S7\!\!]]
    Y.~Gao, S.~A.~Yang, and Q.~Niu,
    \href{https://doi.org/10.1103/PhysRevB.91.214405}{Phys. Rev. B {\bf 91}, 214405 (2015).}

\item[[S8\!\!]]
E.~V.~Gorbar, V.~A.~Miransky, I.~A.~Shovkovy, and P.~O.~Sukhachov,
    \href{https://doi.org/10.1103/PhysRevB.95.205141}{Phys. Rev. B {\bf 95}, 205141 (2017).}

\item[[S9\!\!]]
A.~Cortijo, Y.~Ferreir\'{o}s, K.~Landsteiner, and M.~A.~H.~Vozmediano,
    \href{https://doi.org/10.1103/PhysRevLett.115.177202}{Phys. Rev. Lett. {\bf 115}, 177202 (2015).}

\item[[S10\!\!]]
   A.~Cortijo, D.~Kharzeev, K.~Landsteiner, and M.~A.~H.~Vozmediano,
    \href{https://doi.org/10.1103/PhysRevB.94.241405}{Phys. Rev. B {\bf 94}, 241405(R) (2016).}

\item[[S11\!\!]]
    K.~Landsteiner,
    \href{https://www.actaphys.uj.edu.pl/R/47/12/2617}{Acta Phys. Polon. B {\bf 47}, 2617 (2016).}

\item[[S12\!\!]]
    E.~V.~Gorbar, V.~A.~Miransky, I.~A.~Shovkovy, and P.~O.~Sukhachov,
    \href{https://doi.org/10.1103/PhysRevLett.118.127601}{Phys. Rev. Lett. {\bf 118}, 127601 (2017).}

\item[[S13\!\!]]
Z.~Wang, Y.~Sun, X.-Q.~Chen, C.~Franchini, G.~Xu, H.~Weng, X.~Dai, and Z.~Fang,
    \href{https://doi.org/10.1103/PhysRevB.85.195320}{Phys. Rev. B {\bf 85}, 195320 (2012).}

\item[[S14\!\!]]
Z.~Wang, H.~Weng, Q.~Wu, X.~Dai, and Z.~Fang,
    \href{https://doi.org/10.1103/PhysRevB.88.125427}{Phys. Rev. B {\bf 88}, 125427 (2013).}

\item[[S15\!\!]]
E.~V.~Gorbar, V.~A.~Miransky, I.~A.~Shovkovy, and P.~O.~Sukhachov,
    \href{https://link.aps.org/doi/10.1103/PhysRevB.91.121101}{Phys. Rev. B {\bf 91}, 121101(R) (2015).}

\item[[S16\!\!]]
A.~A.~Abrikosov, {\sl Fundamentals of the Theory of Metals} (North-Holland, 1988).

\item[[S16\!\!]]
C.~Kittel, {\sl Quantum Theory of Solids} (Wiley, 1987).

\item[[S17\!\!]]
V.~V.~Gudkov and J.~D.~Gavenda, {\sl Magnetoacoustic Polarization Phenomena in Solids} (Springer Science and Business Media, 2000).

\item[[S18\!\!]]
W. Freyland, C. Madelung, A. Goltzene, P. Grosse, et al., {\sl Non-Tetrahedrally Bonded Elements and Binary Compounds I, Condensed Matter} (Springer-Verlag Berlin Heidelberg, 1998).

\item[[S19\!\!]]
    H. Wang, Y. Xu, M. Shimono, Y. Tanaka, and M. Yamazaki,
    \href{http://dx.doi.org/10.2320/matertrans.MAW200717}{Mater. Trans. {\bf 48}, 2349 (2007).}

\item[[S20\!\!]]
M.~Neupane, S.-Y.~Xu, R.~Sankar, N.~Alidoust, G.~Bian, C.~Liu, I.~Belopolski, T.-R.~Chang, H.-T.~Jeng, H.~Lin, A.~Bansil, F.~Chou, and M.~Z.~Hasan,
    \href{https://doi.org/10.1038/ncomms4786}{Nat. Commun. {\bf 5}, 3786 (2014).}

\item[[S21\!\!]]
    Z.~K.~Liu, J.~Jiang, B.~Zhou, Z.~J.~Wang, Y.~Zhang, H.~M.~Weng, D.~Prabhakaran, S-K.~Mo, H.~Peng, P.~Dudin, T.~Kim, M.~Hoesch, Z.~Fang, X.~Dai, Z.~X.~Shen, D.~L.~Feng, Z.~Hussain, and Y.~L.~Chen,
    \href{https://doi.org/10.1038/nmat3990}{Nat. Mater. {\bf 13}, 677 (2014).}

\item[[S22\!\!]]
C.-Z.~Li, L.-X.~Wang, H.~Liu, J.~Wang, Z.-M.~Liao, and D.-P.~Yu,
    \href{https://doi.org/10.1038/ncomms10137}{Nat. Commun. {\bf 6}, 10137 (2015).}

\item[[S23\!\!]]
A.~Polkovnikov,
    \href{https://doi.org/10.1016/j.aop.2010.02.006}{Ann. Phys. (N. Y.) {\bf 325}, 1790 (2010).} 

\item[[S24\!\!]]
A.~A.~Soluyanov, D.~Gresch, Z.~Wang, Q.~Wu, M.~Troyer, X.~Dai, and B.~A.~Bernevig,
    \href{https://doi.org/10.1038/nature15768}{Nature {\bf 527}, 495 (2015).}

\item[[S25\!\!]]
M.~M.~Vazifeh and M.~Franz,
    \href{https://doi.org/10.1103/PhysRevLett.111.027201}{Phys. Rev. Lett. {\bf 111}, 027201 (2013).}

\item[[S26\!\!]]
H.~Shapourian, T.~L.~Hughes, and S.~Ryu,
    \href{https://doi.org/10.1103/PhysRevB.92.165131}{Phys. Rev. B {\bf 92}, 165131 (2015).}

\item[[S27\!\!]]
A.~Cortijo and M.~A.~~Zubkov,
    \href{https://doi.org/10.1016/j.aop.2016.01.006}{Ann. Phys. (N. Y.) {\bf 366}, 45 (2016).}

\item[[S28\!\!]]
M.~A.~Zubkov and M.~Lewkowicz,
    \href{https://doi.org/10.1016/j.aop.2018.08.006}{Ann. Phys. (N. Y.) {\bf 399}, 26 (2018).}

\end{itemize}


\begin{thebibliography}{100}
\bibitem{comment1} The acoustogalvanic effect should be also present in other types of WSMs such as parity-symmetry broken and type-II ones as long as strains can be interpreted in terms of axial gauge fields.
\end{thebibliography}
\end{document}